\documentclass[
aps,%
12pt,%
final,%
notitlepage,%
oneside,%
onecolumn,%
nobibnotes,%
nofootinbib,%
superscriptaddress,%
noshowpacs,%
centertags]%
{revtex4}

\input{epsf}
\usepackage[dvipsnames]{color}

\def\ltsima{$\; \buildrel < \over \sim \;$}
\def\simlt{\lower.5ex\hbox{\ltsima}}
\def\gtsima{$\; \buildrel > \over \sim \;$}
\def\simgt{\lower.5ex\hbox{\gtsima}}
\def\gee{ \, \lower 1mm\hbox{$\,{\buildrel > \over{\scriptstyle\scriptstyle\sim} }\displaystyle \,$}}
\def\lee{ \, \lower 1mm\hbox{$\,{\buildrel < \over{\scriptstyle\scriptstyle\sim} }\displaystyle \,$}}
\def\Oo {\displaystyle}

 \addto\captionsrussian{}

\begin{document}
\selectlanguage{russian}

\title{\bf Polygonal Structures in a Gaseous Disk: Numerical Simulations}
\author{\firstname{S.A.}~\surname{Khoperskov}
 } \email{shoperskov@gmail.com}
 \author{\firstname{A.V.}~\surname{Khoperskov}
 } \email{khoperskov@volsu.ru}
  \author{\firstname{M.A.}~\surname{Eremin}
 }
  \author{\firstname{M.A.}~\surname{Butenko}
 }
\affiliation{%
Volgograd State University,\\
Vtoraya Prodolnaya ul. 30, Volgograd, 400062 Russia
}%

\begin{abstract}
The results of numerical simulations of a gaseous disk in the potential of a stellar spiral density
wave are presented. The conditions under which straightened spiral arm segments (rows) form in the
gas component are studied. These features of the spiral structure were identified in a series of works by
A.D. Chernin with coauthors. Gas-dynamic simulations have been performed for a wide range of model
parameters: the pitch angle of the spiral pattern, the amplitude of the stellar spiral density wave, the disk
rotation speed, and the temperature of the gas component. The results of 2D- and 3D-disk simulations
are compared. The rows in the numerical simulations are shown to be an essentially nonstationary
phenomenon. A statistical analysis of the distribution of geometric parameters for spiral patterns with rows
in the observed galaxies and the constructed hydrodynamic models shows good agreement. In particular,
the numerical simulations and observations of galaxies give $\langle{\alpha}\rangle\simeq 120^\circ$
for the average angles between straight segments.

\end{abstract}

\maketitle

%----------------------- Section 1 -------------------------------

\section{PECULIARITIES OF OBSERVED SPIRAL
PATTERNS}

\noindent

\textbf{The Geometric Shape of Spiral Structures}.
 Smooth regular spiral arms in grand-design galaxies
are encountered very rarely. As a rule, the global
spiral pattern is characterized by a large number
of various inhomogeneities on small and medium
spatial scales: spurs (Elmegreen 1980; Chakrabarti
et al. 2003; Shetty and Ostriker 2006; Muraoka
et al. 2009), feathers (Dobbs and Bonnell 2006; La
Vigne et al. 2006), branchings and branch thickenings,
and quasi-periodic complexes along the arms
(Efremov and Chernin 2003; Efremov 2009). Interestingly,
the small-scale inhomogeneity along the
azimuth angle is also pronounced in ring structures,
such as NGC 7742 and NGC 7217 (Silchenko
and Moiseev 2006). The so-called rows or polygonal
structures (PSs) by which we will mean the
sequences of extended almost straight segments
forming the spiral patterns of many galaxies (Waller
et al. 1997; Pohlen and Trujillo 2006) are larger-scale
features of the spiral pattern. Vorontsov-Vel▓yaminov
was the first to notice such objects, calling them
rows (Vorontsov-Vel▓yaminov 1964, 1977). In a
series of works, Chernin with coauthors considered
in detail the properties of PSs for nearby galaxies
(Chernin 1998, 1999a; Chernin et al. 2000, 2001a)
and our Galaxy (Chernin 1999b). There is evidence
for the presence of straightened segments of the
spital structure in M31 (Efremov 2001, 2009, 2010).
The so-called <<hexagonal>> shape in some internal
ring structures of galaxies may be similar in nature
(Chernin et al. 2001b).

Polygonal structures are revealed not only in the
optical range by young star clusters by also by the
distribution of interstellar $HI$ gas, dust, synchrotron
emission, ultraviolet emission, and $H_\alpha$. Meanwhile,
the spiral arms formed by old stars appear smoother
in infrared images of galaxies. Figure 1 schematically
shows the PSs that we identified for a number of
observed objects additional to the catalog by Chernin
et al. (2001a).

\begin{figure}[!t]
 \setcaptionmargin{5mm} \onelinecaptionstrue
\includegraphics[width=0.92\hsize]{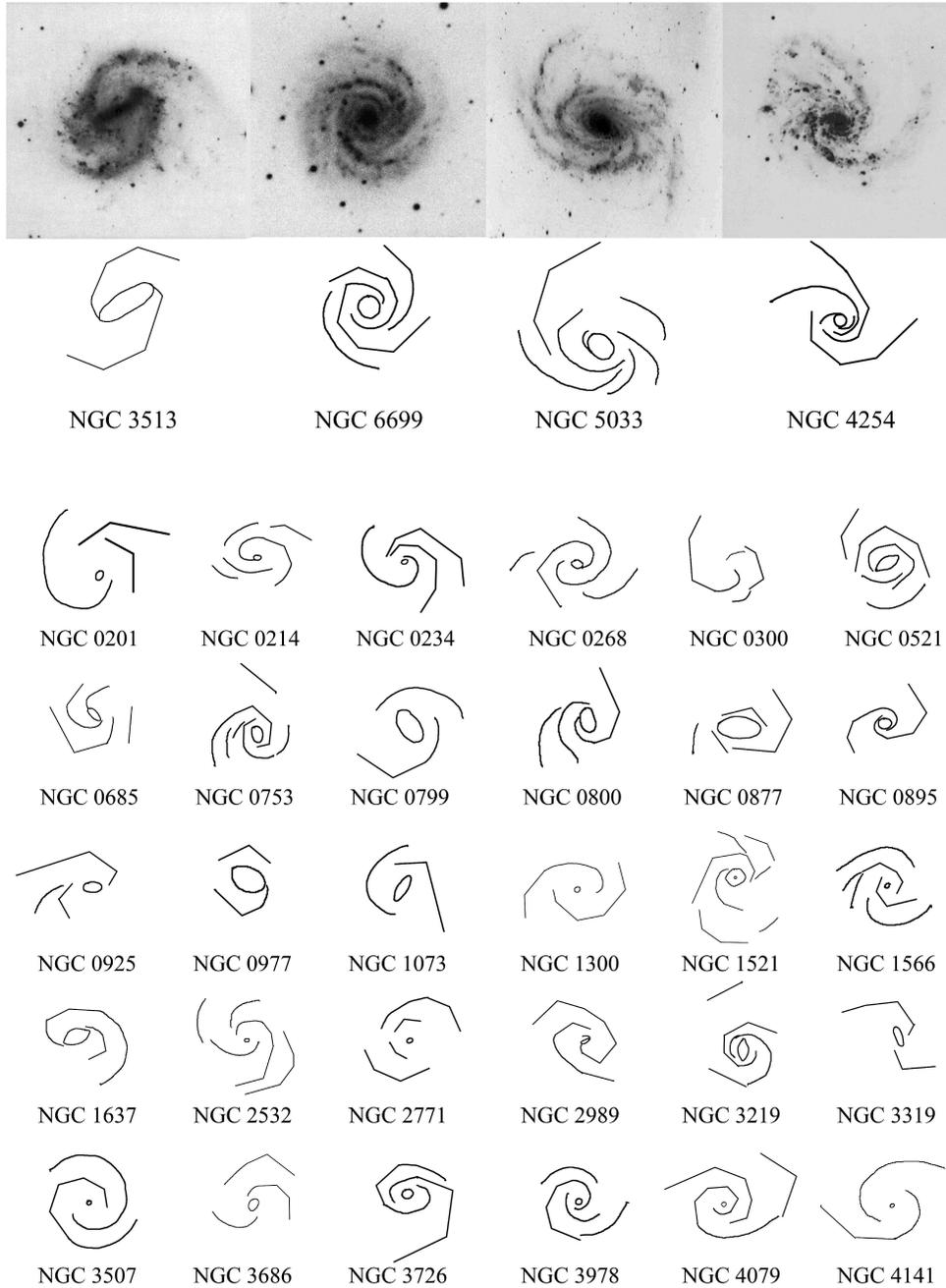}
\captionstyle{normal}
 \vskip -0.12\hsize \vbox{\hsize=0.9\hsize  \caption{New examples of the schematic view of galaxies with PSs additional to the data from Chernin et al. (2001a).
\hfill} }\label{fig01}
\end{figure}

Note the most characteristic features of polygonal
structures from observational data (see Chernin
1998, 1999Ю, 1999b; Chernin et al. 2000, 2001a).
\begin{enumerate}
   \item For the sample of nearest galaxies constructed
by Chernin et al. (2001a), the number of objects with
PSs exceeds 200, which accounts for about 5--10\%
of the total number of galaxies with a global spiral
structure.

The average angle between neighboring
straight segments from observational data is \mbox{$\langle\alpha\rangle\simeq 120^\circ$} (Chernin et al. 2000). The scatter of values
lies within the range  \mbox{$100^\circ \lee \alpha \lee 145^\circ$} (Chernin
et al. 2001a). The length of such segments $\ell$,
on average, increases in direct proportion to the galactocentric distance $d$ (Chernin 1999a; Chernin
et al. 2001a).

  \item PSs are present both in SB galaxies and in
systems without any central bar, but mostly in latertype
objects (Sbc, Scd).

  \item The relative abundance of gas is significant,
$M_{HI}/L \simeq 0.15 - 0.4$ ($M_{HI}/L$ is the mass-to-light ratio
in solar units), but exceptions are possible, for
example, this ratio for NGC~4548 is only \mbox{$M_{HI}/L\simeq 0.04$}
(Chernin et al. 2000).

  \item The rotation speed estimates for the galaxies
considered are not very reliable, because the disks are
oriented almost face-on, but, on the whole, the maximum
rotation speeds (in the plateau region) exceed
150 $\text{km s}^{-1}$.

  \item An enhanced fraction of interacting galaxies is
observed among objects with PSs. The longest rows
($\ell \gtrsim 8$\,kpc) are encountered mostly in interacting
galaxies, while shorter rows ($\ell \lesssim 5$\,kpc) are observed
a factor of $\simeq 4$ more often in galaxies without any
apparent manifestations of interaction.

  \item  The number of rows identified in various objects
lies within the range $n_{row}=1\div 9$ and the mean
number in the sample is 3 rows.

  \item Rows (kinks and straight segments) can
be identified in some ring galaxies; Chernin et al.
(2001c) called such systems hexagonal structures
(e.g.,NGC\,7020, NGC\,4429, NGC\,3351, NGC\,6782,
NGC\,6935, UGC\,12646, ESO 325-28) (see also
Buta and Combes 1996)).
\end{enumerate}
Apart from galaxieswith a large number of straight
segments of spirals for both spiral arms, in several
galaxies kinks can be identified only in one of the
galactic arms (see NGC\,4548 and NGC\,7137,
Chernin et al. 2000).

Several mechanisms have been proposed to explain
the formation of straightened segments of
galactic spiral arms. According to the hypothesis
suggested by Contopoulous and Grosbol (1986), the
presence of PSs is associated with the influence of
resonance regions in the stellar population dynamics.
Chernin (1999a) and Chernin et al. (2000) proposed
an alternative mechanism for the formation of polygonal
structures: the straightened segments of spirals
result from the development of instability of the galactic
shock front that leads to its fragmentation into
flat areas. Indeed, based on 2D hydrodynamic simulations
of a nonrotating gas, Chernin et al. (2006)
showed that when the gas flows through a curved
potential well, the front of a plane shock in the case
of oblique onflow breaks up into individual fragments
and the effect of shock escape from the potential well
takes place.

Thus, the observed straightening of spiral segments
in galaxies can be hydrodynamic in nature.
Nevertheless, a number of physically important factors
remained outside the scope of the numerical
simulations described by Chernin et al. (2006), where
the conclusion reached in models without rotation
that the shock front is located at the flow entrance
into the potential well of a stellar density wave seems
to be themain shortcoming. Observational data point
to the positions of the characteristic dust lanes associated
with the shock opposite to those in Chernin
et al. (2006). This manifests itself particularly clearly
in barred galaxies when the corotation radius is
near the ends of the stellar bar (see the images of
NGC\,1097, NGC\,1300, NGC\,1365). The dust lane
lies at the leading edge inside the bar and at the
trailing edge in the spirals outside the bar.

\textbf{Barred Galaxies}.
Traditionally, the spiral pattern in SB galaxies is
associated with a nonaxisymmetric perturbation of
the central stellar bar (or oval). Until recently, under
the assumption that the system is stationary, the
kinematics of the spiral structure was believed to be
determined by the rotation of a rigidly rotating bar
with a pattern speed $\Omega_{bar}$. If the corotation radius of
the bar $r_{c}^{bar}$ is near its ends, then the corotation of spirals
is also nearby, $r_{c}^{spir}\simeq r_{c}^{bar}$. Thus, the solid-body
rotation speed of the spiral pattern $\Omega_{spir}$ is everywhere
larger than the disk rotation speed  $\Omega(r)=V(r)/r<\Omega_{spir}$. Let us separately discuss the question about
PSs in galaxies with a central bar/oval based on the
OSUBGS sample of galaxies for which the corotation
radii $r_c$ were determined in a number of works (Rautiainen
et al. 2008; Buta and Zhang 2009). A characteristic
feature of the analysis performed by Buta and
Zhang (2009) is the existence of an outer corotation
radius rc outside the apparent spiral structure ${\cal R}=r_c/r_{bar}\gee 2$ for more than 60\% of the galaxies. This
points to slow rotation of at least the outer spiral arms,
$\Omega_{spir}< \Omega(r)=V/r$.

The estimated positions of the relative corotation
radius ${\cal R}=r_c/r_{bar}$ for 38 SB galaxies from the OSUBGS
sample in Rautiainen et al. (2008) show that
${\cal R}=r_c/r_{bar}$ exceeds appreciably unity for a sizeable
fraction of the galaxies. We will point out NGC\,0289,
NGC\,0578, NGC\,1187, NGC\,3726, NGC\,4051,
NGC 4995, and NGC 6384 with ${\cal R}=r_c/r_{bar}=1.8 - 3.4$. Other examples of such objects are also known:  ${\cal R}_{NGC\,3081}=2.2$
(Buta and Purcell 1998) and ${\cal R}_{NGC\,0925}=3.1$ (Elmegreen et al. 1998). Based on the results by Zhang and Buta (2007, 2010), we
can identify the galaxies NGC\,5194, NGC\,5247,
NGC\,4321, NGC\,4622, and NGC\,1073 for which
the outer corotation radius is outside the spiral
structure or on its periphery. Analysis of the images
for OSUBGS galaxies with ${\cal R}\gee 2$ from the papers
listed above shows that NGC\,1187, NGC\,1385,
NGC\,3513, NGC\,3686, NGC\,4145, NGC\,4902,
NGC\,5194, NGC\,5334, and NGC\,7418 have rows.
The existence of several corotation radii is an argument
for the necessity of changing the common
viewpoint on invariability of the morphological type
and stationarity of the spiral pattern in the extreme
case of a very slow change in the parameters of spirals
in favor of faster transformations with characteristic
times of the order of the revolution period.

Several corotation radii may also simultaneously
exist in our Galaxy, which is indicative of nonstationary
dynamics (Gerhard 2010, and references therein).
Gerhard (2010) again draws attention to the old
contradiction between a rapidly rotating bar with a
pattern speed $\Omega_{bar} \sim 59$\,$\text{km s}^{-1}\text{kpc}^{-1}$ and a slowly
rotating spiral pattern in the solar neighborhood with
$\Omega_{sp}\simeq 25$\,$\text{km s}^{-1}\text{kpc}^{-1}$. This suggests that there is
no dynamical connection between the central bar and
the spiral structure, at least on the periphery of our
Galaxy.

Note yet another approach to determining the
corotation radius associated with the presumed correlation
between the peculiarity of the radial metallicity distribution $Z(r)$ in the Galactic disk (a local minimum and a plateau) and the position of
the corotation radius, which follows, in particular,
from chemical-dynamical models for the evolution
of galaxies (Mishurov et al. 2002). For our Galaxy,
in addition to the pattern speed of the rapidly rotating
bar, the profile $Z(r)$ points to a corotation
radius $r_c\sim R_\odot$ and a spiral pattern speed $\Omega_{sp}\sim
 25\div 35$\,$\text{km s}^{-1} \text{kpc}^{-1}$ (Mishurov et al. 2002; Acharova
et al. 2010). Observational data for open star
clusters, given their age, suggest that the corotation
radius is close to the position of the solar orbit in
the Galaxy, $r_c/R_\odot = 1.06\pm 0.08$ (Dias and Lepine
2005). This approach allows the position of the
corotation radius on the periphery of the stellar disk to
be determined from observational data, for example,
$r_c/R_{opt}=0.83$, for IC\,0167, $r_c/R_{opt}^{NGC\,1042}=0.65$, $r_c/R_{opt}^{NGC\,6907}=1.0$ (Scarano et al. 2011); all three
galaxies are late-type ones with a bar, with the
corotation radius being at its ends. Note also that
the approaches by Scarano et al. (2011) and Buta
and Zhang (2009) give close values for NGC\,1042, $r_c/R_{opt}^{NGC\,1042}=0.65$ and 0.6, respectively.
The observations of young objects in the vicinity of a spiral
arm (Grosbol and Dottori 2009) point to the position
of the corotation radius $r_c=240''$ near the optical
radius $R_{opt}=270''$ for NGC\,2997; a well-defined
system of rows can be seen in this galaxy.

The rows starting directly from one bar end that
form a ringlike spiral going to the other bar end are
a peculiarity of several barred galaxies. NGC\,1097,
NGC\,2523, and NGC\,4902 are typical representatives.
In numerical N-body models, separate
straightened arm segments can be seen at the transient
formation stage of the bar and spirals (Buta and
Combes 1996).

Note also the theoretical spiral structure models
indicative of a peripheral position of the corotation radius
(Contopoulous and Grosbol 1986). Suchmodels
based on the 4/1 resonance allow the observed peculiarities
of spiral galaxies to be described (Patsis
et al. 1997). As another example, we can point to
the dynamics of a galactic disk in a nonaxisymmetric
massive dark halo (Khoperskov et al. 2007). A slowly
rotating two-armed spiral wave with the corotation
located in the outer disk is generated in such models.
The conclusion that the corotation radius is on
the periphery of the spiral pattern in an appreciable
number of galaxies is important for the results of
our numerical simulations obtained here, because the
galactic shock is straightened most easily in the case
of a slowly rotating spiral pattern where the corotation
is on the disk periphery.

\textbf{Sample of Galaxies with Rows}. In addition to the catalog of 200 galaxies with rows published by Chernin et al. (2001Ю), we identified
102 more spiral galaxies with typical PS fragments
traceable in their images (see Fig. 1). We used data
from LEDA, NED, Pohlen and Trujillo (2006), and
SDSS to construct the sample.

In a number of galaxies, the rows can be identified
even by old stars (e.g., based on 2MASS data):
NGC\,2523 and NGC\,5653. Such objects are rare,
but polygonality in them is revealed by several components.

% NGC 4319. цЕНЛЕРПХЪ ОШКЕБНИ ЙНЛОНМЕМРШ ЯСЫЕЯРБЕММН ПЮГМХРЯЪ, НАПЮГСЪ НВЕМЭ ОПНРЪФЕММШИ БШОПЪЛКЕММШИ ТПЮЦЛЕМР, ОЕПЕЯЕЙЮЪ НОРХВЕЯЙХИ(?) ПСЙЮБ.

The goal of this paper is to investigate the possibility
of PS generation in the gaseous disks of spiral
galaxies by taking into account the following factors:
\begin{description}
    \item[1)] the gravitational potential of a radially inhomogeneous spiral density wave in the stellar disk;
    \item[2)] the solid-body rotation of the spiral stellar disk
structure;
    \item[3)] the differentiality of the gas rotation;
    \item[4)] the radial inhomogeneity of the gaseous disk
parameters;
    \item[5)] the vertical gas motions.
\end{description}
 The listed factors are new compared to those investigated
by Chernin et al. (2006) and they allow the
most significant observed morphological features of
rows in the spiral patterns of several galaxies to be
reproduced.

%----------------------- Section 2 -------------------------------

\section{GASDYNAMIC MODEL}

\noindent

The flow of gas in the external potential of a stellar
disk and a dark halo is described by the equations of
classical gas dynamics. We will assume that the gas
is ideal and polytropic with an adiabatic index close
to unity, $1< \gamma < 1.1$. This ensures an efficient allowance
for cooling within the framework of a singlecomponent
model. We neglect the gas self-gravity
effects. The system of gas-dynamic equations can
then be written as
\begin{eqnarray}
 \frac{\partial \varrho}{\partial t} + \nabla \cdot (\varrho \textbf{u}) &=& 0, \label{gasdyn1}\\
  \frac{\partial \varrho \textbf{u}}{\partial t} + \nabla \cdot (\varrho \textbf{u}\otimes\textbf{u} ) &=&  -\nabla{p} -\varrho\nabla\Psi, \label{gasdyn2}\\
   \frac{\partial E}{\partial t} + \nabla \cdot ( [E + {p}]\textbf{u} ) & =& -\varrho \textbf{u} \cdot \nabla\Psi,\label{gasdyn3}
\end{eqnarray}
where $\varrho$ is the gas volume density, ${p}$ is the pressure,
$\textbf{u}=\{u,v,w\}$ is the gas velocity vector, and $\Psi$ is the
external gravitational potential. The bulk energy and
the internal specific energy are defined, respectively,
by the expressions
$\Oo E = \varrho
\left(e + \frac{\textbf{u}^2}{2}\right)$,  \mbox{$\Oo e =
\frac{p}{\varrho\,(\gamma - 1)}$}.

Following Wada and Koda (2001), Cox and
Gomez (2002), Shetty and Ostriker (2006), we will
represent the external potential as the sum of two
parts: an axisymmetric one $\Psi_0$ attributable to the
halo and an axisymmetric distribution of matter in
the stellar disk and a nonaxisymmetric one associated
with the density wave in the stellar disk. It is then
convenient to write the potential as
\begin{equation}\label{Pot}
\Psi(r,\varphi,z) = \Psi_0(r,z)\cdot
[1+\varepsilon_0\Psi_1(\xi_s)\cos \Theta_p]\,, \quad \Psi_1=\frac
{\xi_s^2}{(1+\xi_s^2)^{3/2}}\,,
\end{equation}
where
\begin{equation}\label{PotPar}
\xi_s=\sqrt{(r/b)^2+(z/h_\ast)^2}, \quad \Oo\Theta_p =
m\left[\varphi - \varphi_p(r_0) + \Omega_p t
-\frac{\ln(r/r_0)}{\mathrm{tg}\,i}\right]\,,
\end{equation}
$b$, $h_\ast$ are the radial and vertical scale lengths,
respectively, $m$ is the number of spirals, $i$ is the spiral
pitch angle, $\Omega_p$  is the spiral pattern speed, and $\varepsilon_0$ characterizes the depth of the potential well for the spiral density wave. In our calculations, we took
the following dimensionless parameters:  $b=1$, $h_\ast=0.1$, $r_0=0.9$. For the exponential scale length of an
axisymmetric stellar disk with a radial surface density
profile $\sigma_\ast\propto \exp(-r/L_\ast)$, we assumed that $L_\ast=0.5$. Thus, the optical radius of the model galaxy is $R_{opt}\simeq 4L_\ast = 2$.

For this choice of dimensionless parameters,
the disk revolution period at radius $R_{opt}$ is $\tau(r=2)\simeq 6$. The potential of a quasi-isothermal halo provides
a plateau-type curve far from the disk center.
We will assume that the relative mass of the dark halo
within the optical radius is a factor of $1.5\div 3$ larger
than the mass of the disk subsystem (stars + gas)
(Zasov et al. 2004; Khoperskov et al. 2010).

The system of equations (\ref{gasdyn1}) -- (\ref{gasdyn3}) was integrated numerically in a reference frame rotating with the spiral
pattern speed $\Omega_p$. For our computer simulations
of gaseous disks, we implemented a finite-volume
numerical TVD MUSCL scheme (van Leer 1979;
Harten 1983) of the second order in time and the
third order in space inCartesian, polar, and cylindrical
coordinate systems. The TVD method is efficient for
describing an essentially nonstationary flow in which
a system of small-scale shocks is formed (Khoperskov
et al. 2003). The size of the computational
region in 2D models reached $15\times 15$ in the Cartesian
coordinate system, while the outer boundary in the
polar coordinate system was located at radius $r_{\max}=7$. To eliminate the numerical boundary effects in
the fictitious cells of the computational region, we
assumed the gas parameters to be equal to the analytical
ones or the gaseous disk to be immersed into
a vacuum over the entire computation time. To use
the latter alternative, we developed and implemented
an approach that allowed the evolution of the matter√
vacuum boundary to be computed in numerical models
(Eremin et al. 2010).

At the initial time, we assumed that the gas surface
density was in the form of a power law:
\begin{equation}
\sigma(r, t=0) = \frac{1}{[1+(r/L_\sigma)^2]^{5/2}}\,,
\end{equation}
where $L_\sigma\simeq 1\div2$ is the spatial scale, and that the
isentropic condition was met:
\begin{equation}
p(r,z,t=0) = K \cdot \varrho^\gamma(r,z,t=0), \quad
K=\mathrm{const}\,.
\end{equation}
The radial profiles of the flow parameters in the equatorial
$z=0$ plane were determined from the conditions
for the gaseous disk being axisymmetric and
equilibrium for  $\varepsilon_0=0$. For 3D models, the disk at
the initial time was assumed to be also in a state of
hydrostatic equilibrium in the vertical direction.

The numerical grids are determined by the number
of cells in the radial, $N_r$, azimuthal, $N_\varphi$, and vertical,
$N_z$, directions. For our 2D simulations, we used
several groups of models: $N_r=500\div 3200$, $N_\varphi=360\div 1080$. In our 3D simulations, the parameters of
the numerical grid are $N_r=1000$, $N_\varphi=360$, $N_z=400$. In the best models as applied to a typical galaxy
with $R_{opt}=10$\,kpc, the corresponding spatial resolution
reaches $\sim 10$~pc and the angular resolution in
azimuth reaches $\Delta\varphi = (1/3)^\circ$.

The main parameters of our numerical models are
the amplitude $\varepsilon_0$ responsible for the depth of the spiral
potential well, the spiral pitch angle~$i$, the spiral pattern
speed $\Omega_p$, and the speed of sound $c_s\simeq \sqrt{K}$ in the
plateau region of the rotation curve. It is convenient
to use the effective Mach number${\cal
M}_0=V_{\max}/c_s^{\max}$ to characterize the initial temperature in the disk.

\begin{figure}[!t]
 \setcaptionmargin{5mm} \onelinecaptionstrue
\begin{tabular}{ccc}
 \includegraphics[width=0.35\hsize]{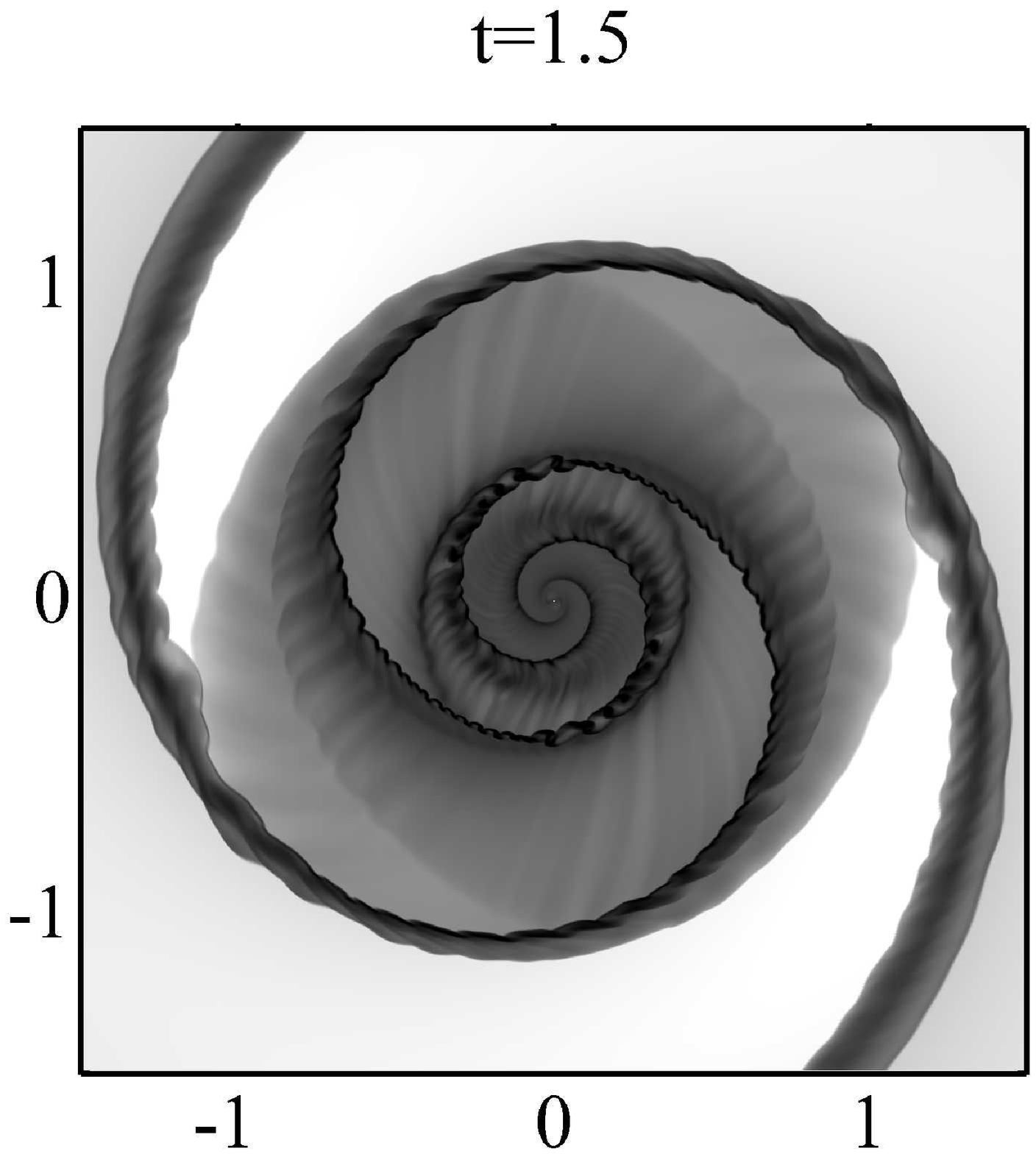}  &
 \includegraphics[width=0.35\hsize]{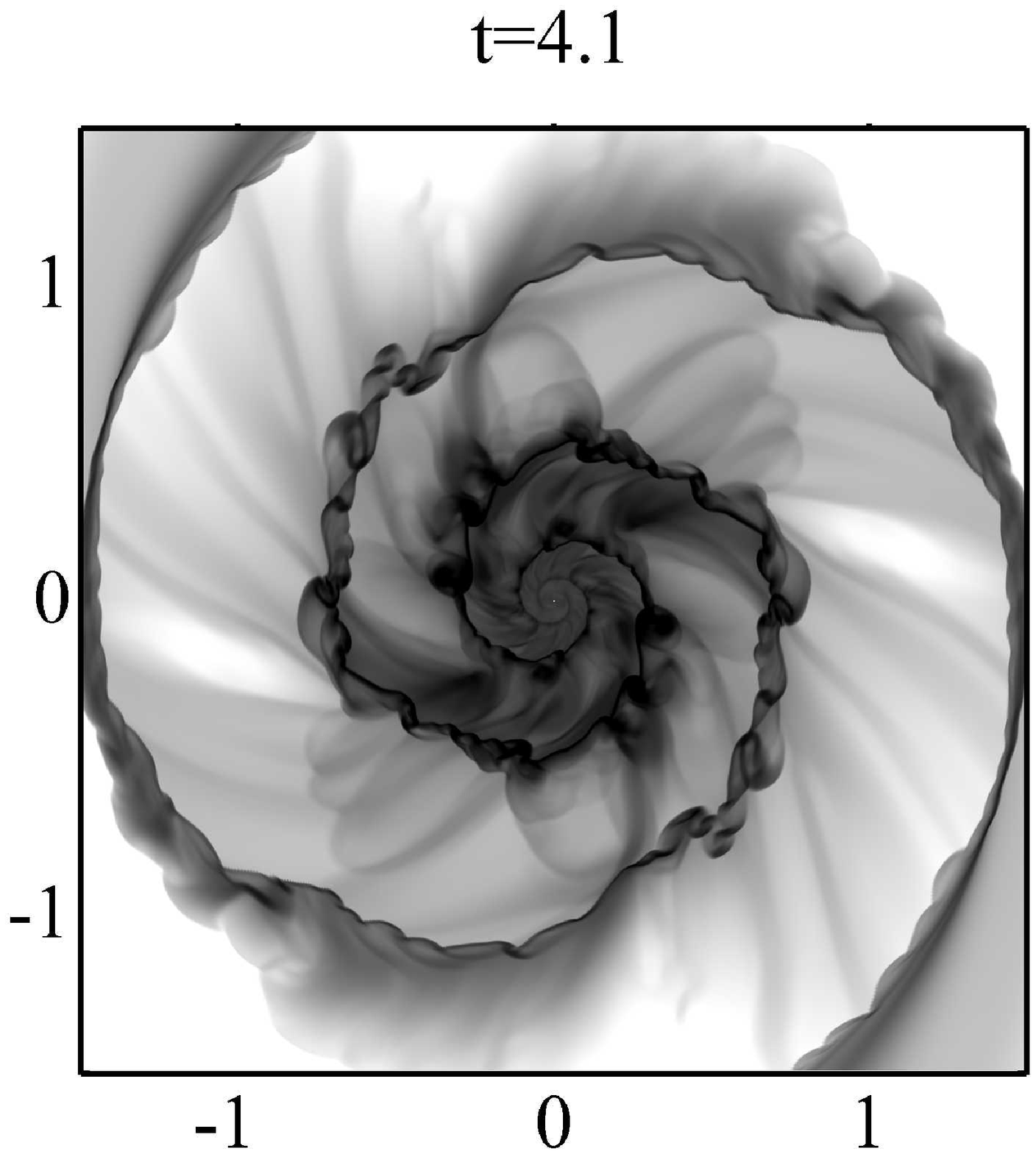} &
 \includegraphics[width=0.35\hsize]{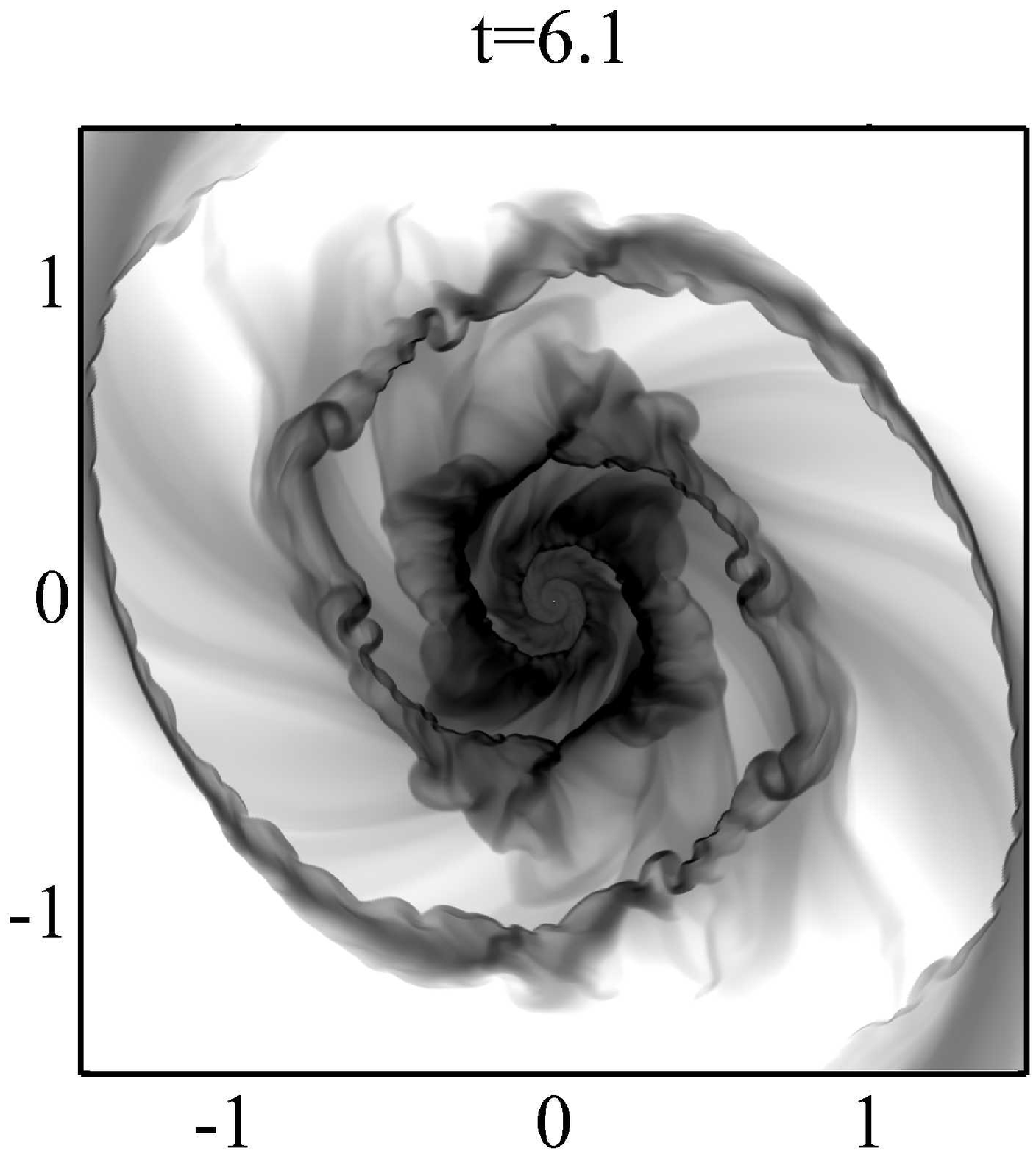} \\
  \includegraphics[width=0.35\hsize]{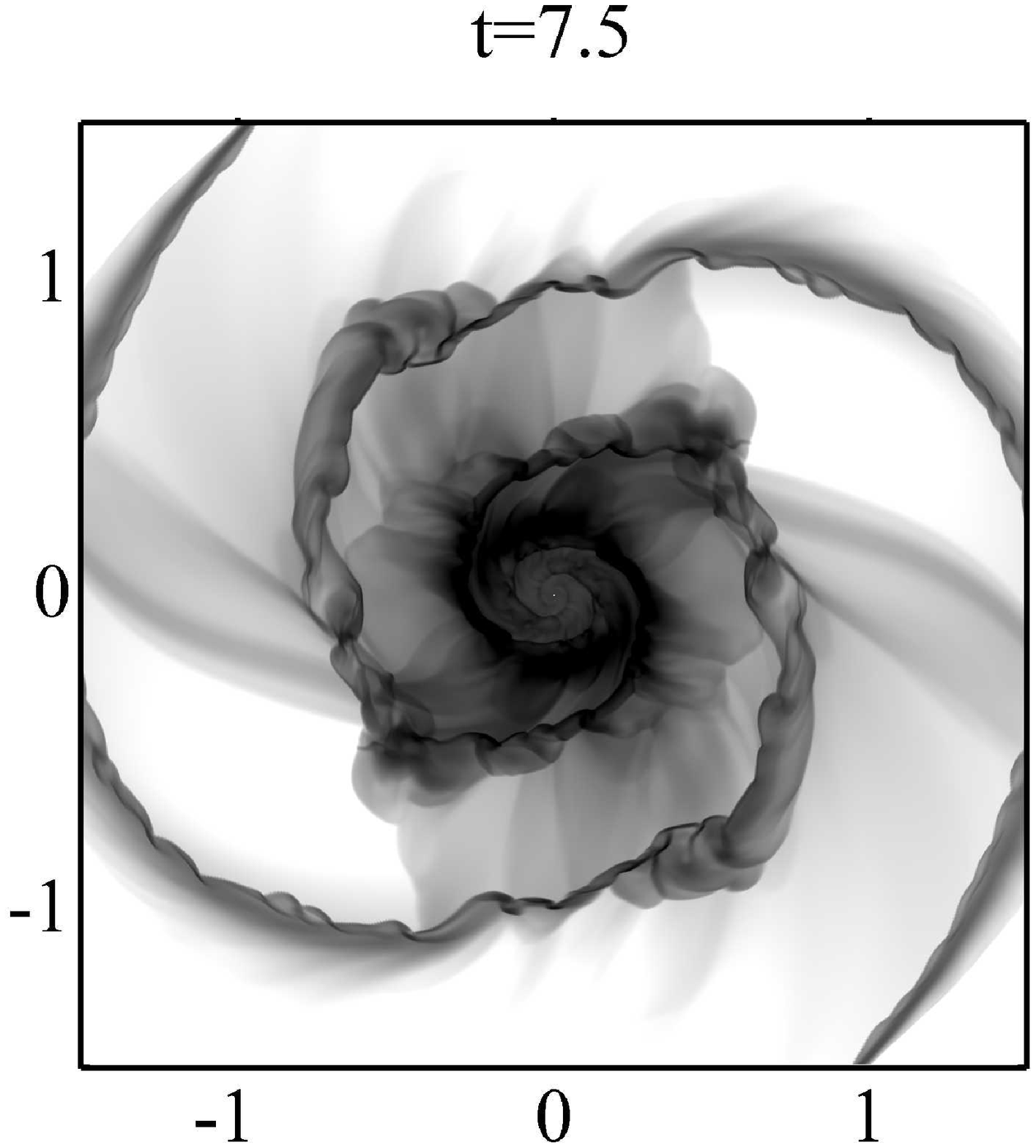} &
  \includegraphics[width=0.35\hsize]{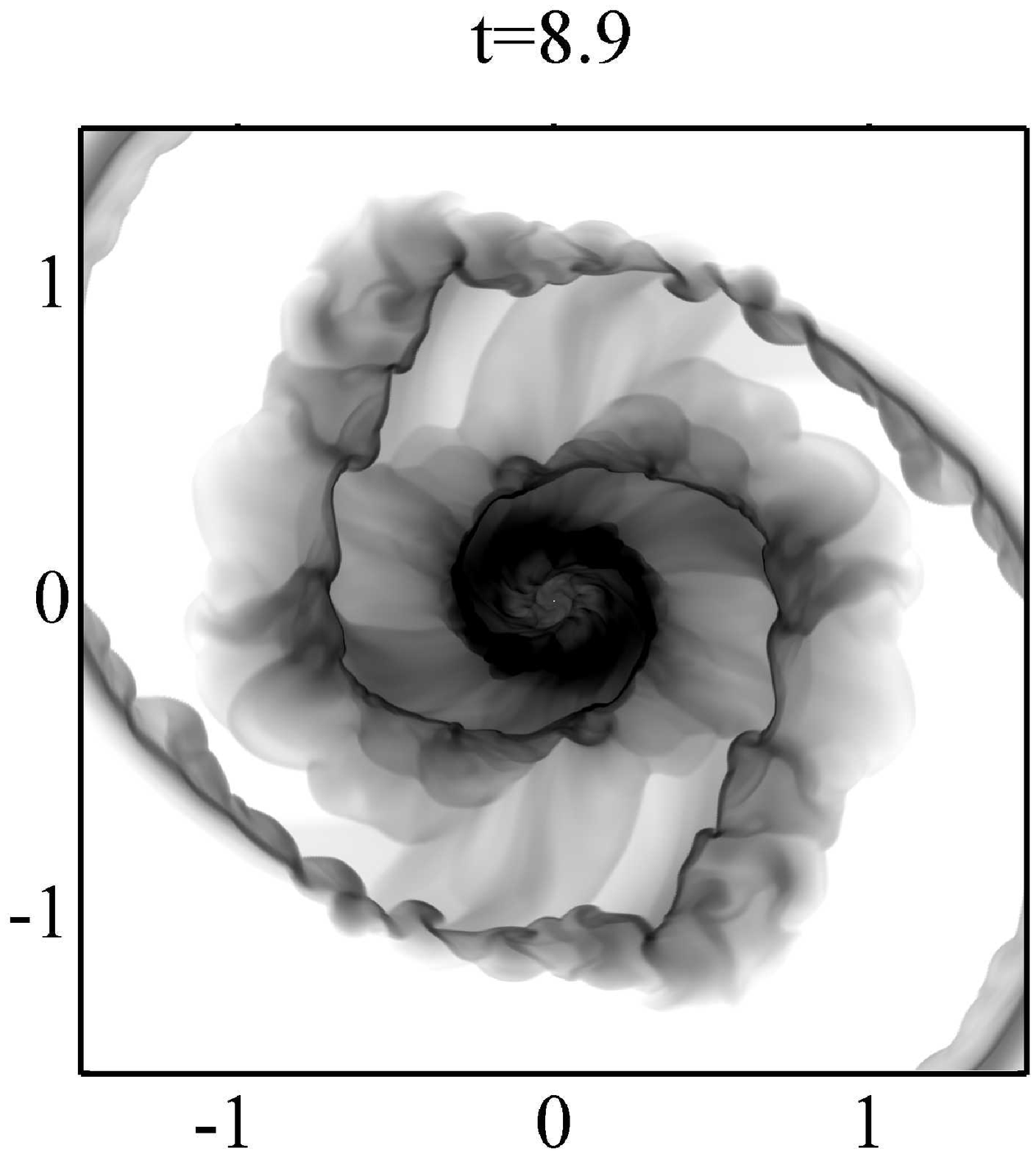} &
  \includegraphics[width=0.35\hsize]{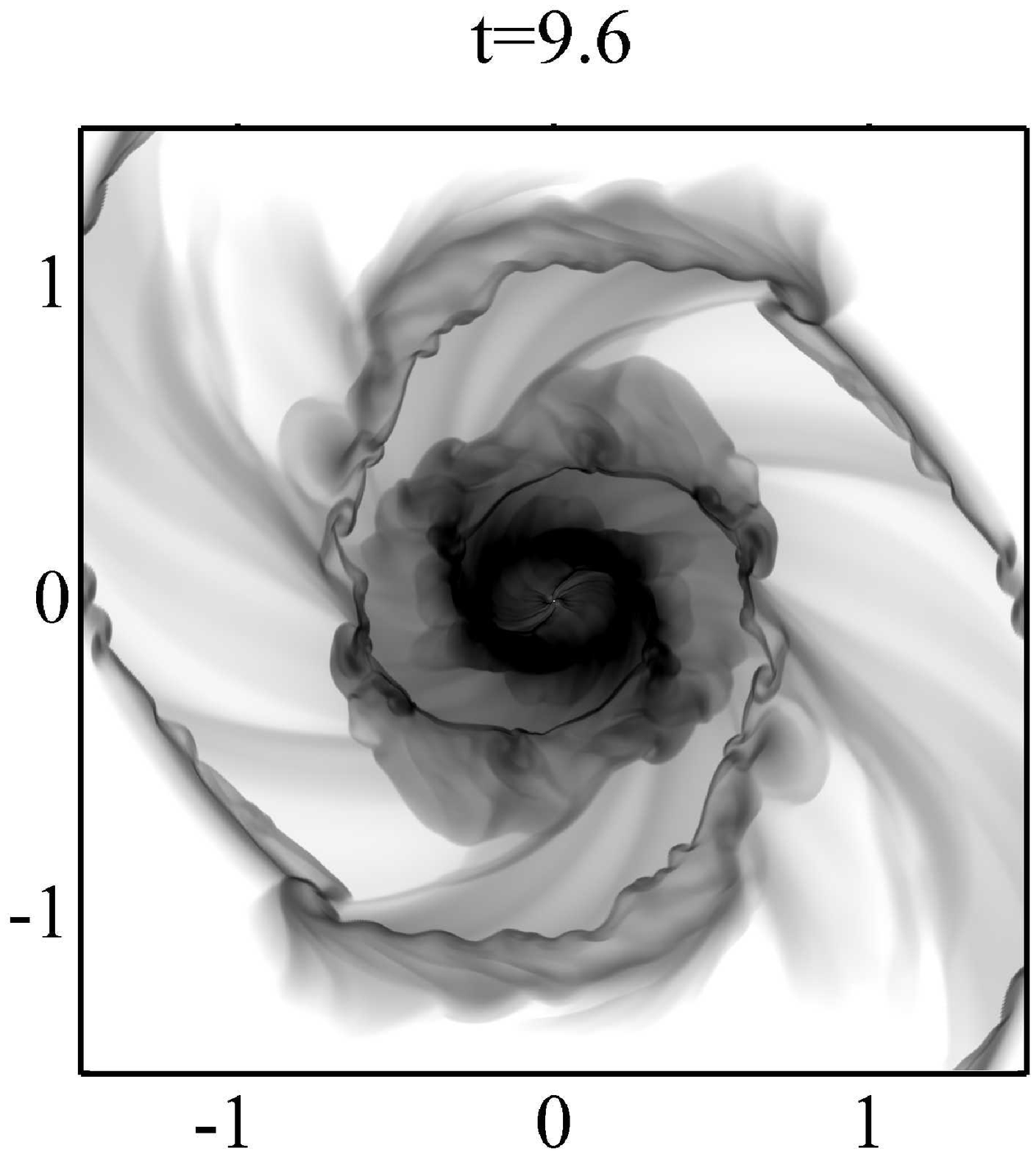} \\
  \includegraphics[width=0.35\hsize]{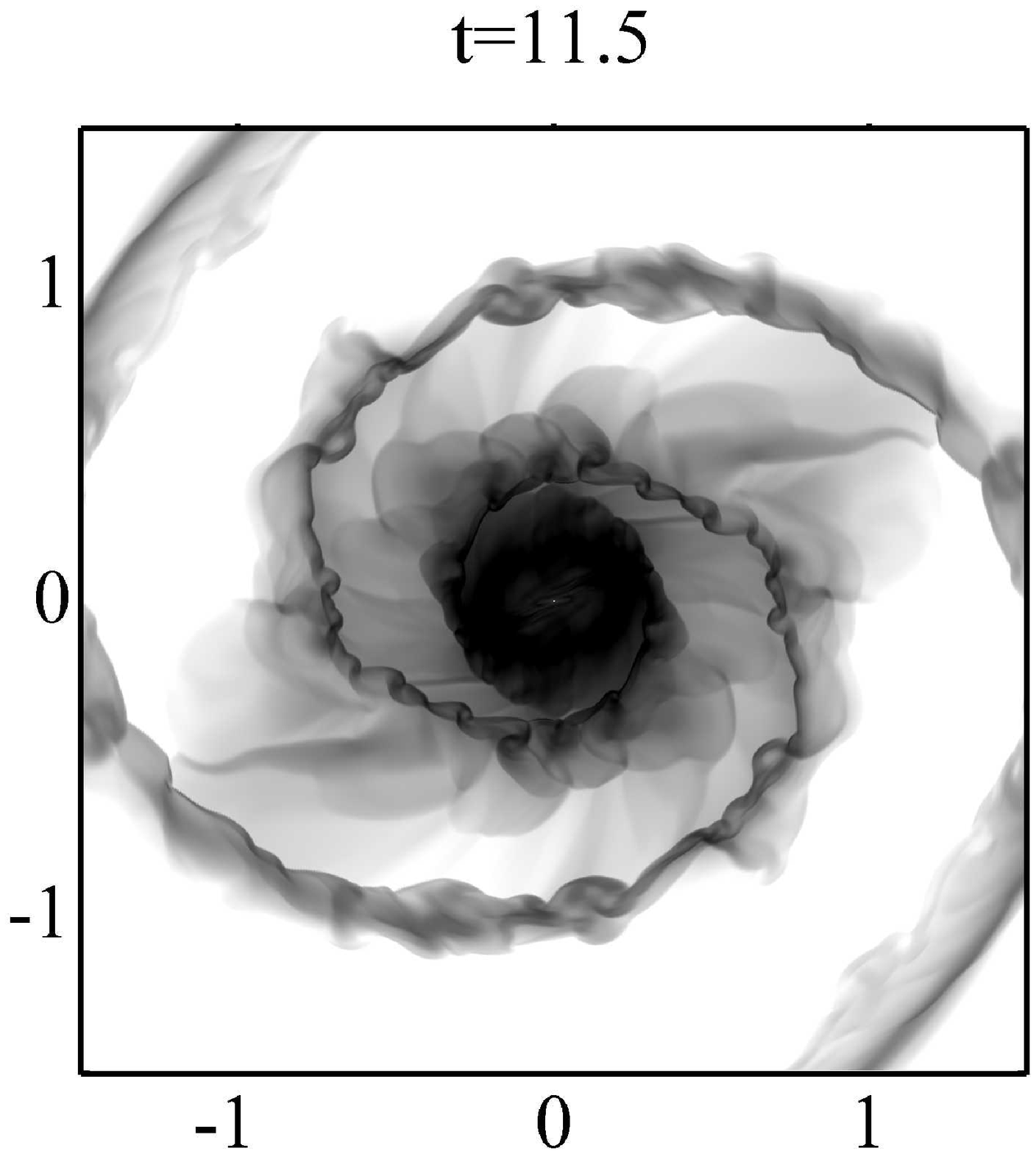} &
  \includegraphics[width=0.35\hsize]{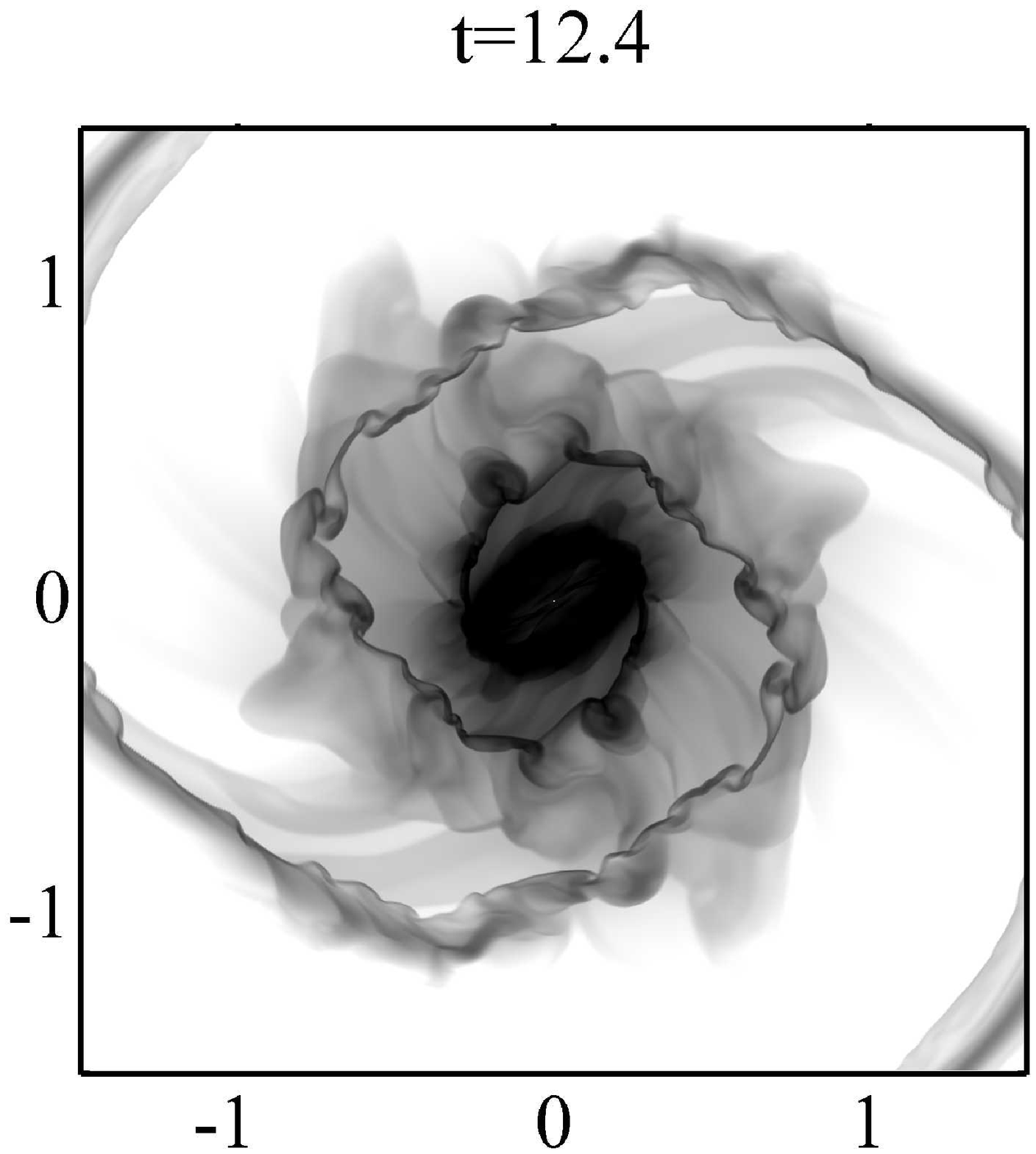} &
  \includegraphics[width=0.35\hsize]{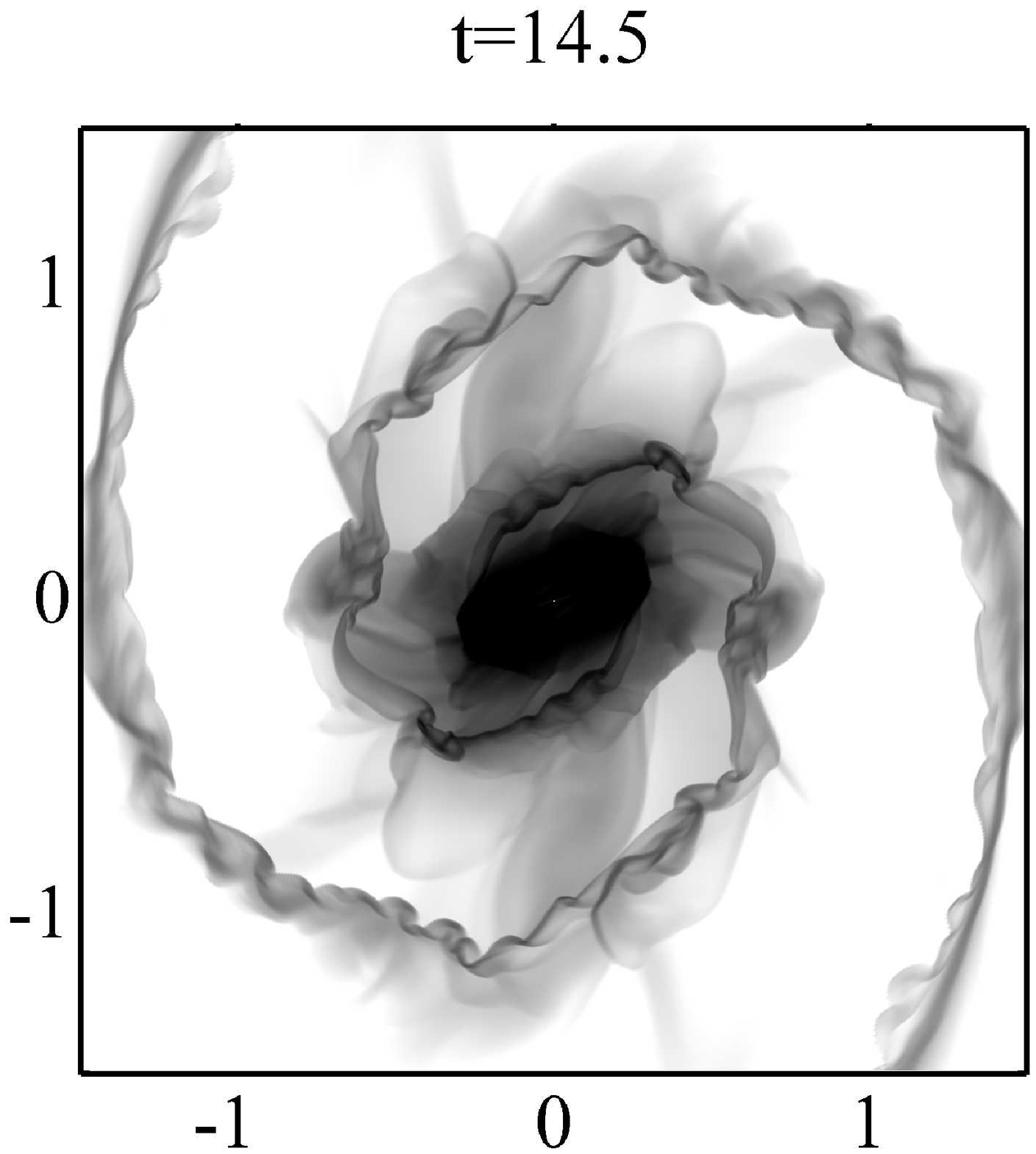}
\end{tabular}
 \captionstyle{normal}
 \vskip -0.02\hsize \vbox{\hsize=0.9\hsize  \caption{Evolution of the gaseous disk in the model with $i=15^\circ$ and $\varepsilon_0=0.15$.
The logarithm of the surface density at various instants of time~$t$. \hfill} }\label{fig02}
\end{figure}

%----------------------- Section 3 -------------------------------

\section{THE FORMATION OF POLYGONAL STRUCTURES}

\noindent

  Our numerical simulations reveal several stages
of gaseous disk evolution in the course of which the
formation of PSs is observed (Fig. 2).

\begin{figure}[!t]
 \setcaptionmargin{5mm} \onelinecaptionstrue
 \vskip 0.\hsize \hskip 0.0\hsize
            \includegraphics[width=1.0\hsize]{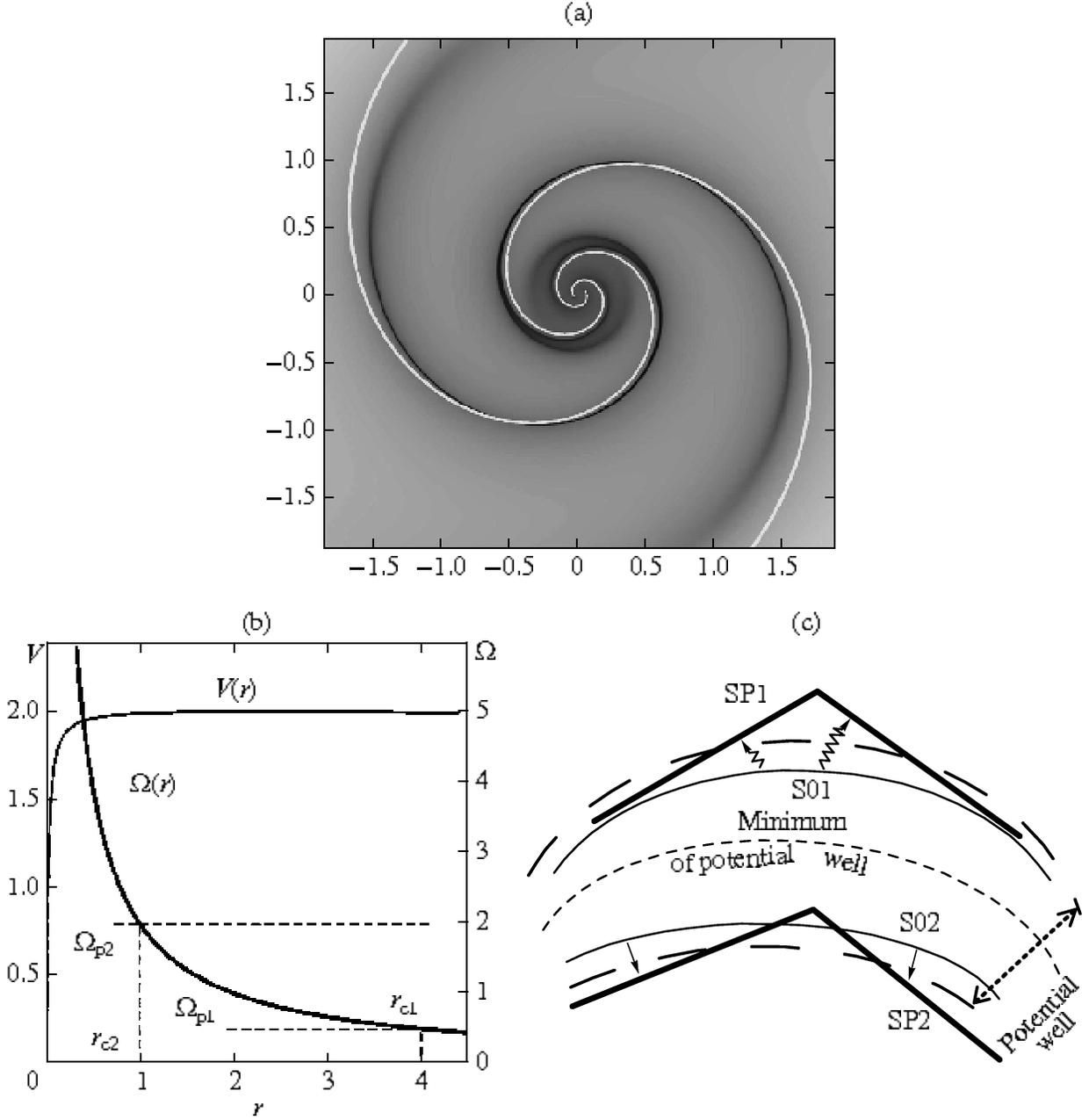}
\captionstyle{normal}
 \vskip -0.\hsize \vbox{\hsize=0.95\hsize  \caption{(a) Position of the shock (the darkest color) in the potential well of a spiral density wave (the white line specifies the
position of the minimum of the potential well for potential (\ref{Pot})). The corotation radius
$r_c=1$. (b) Radial dependences of the
rotation velocity $V(r)$ and angular velocity  $\Omega(r)$ for the gas. For the angular velocity $\Omega_{p1}$, the corotation is on the disk periphery $(r_c=4)$. The case of $\Omega_{p2}$ and $r_{c2}$ corresponds to Fig. 3a. (c) The scheme of shock positions in the potential well of a spiral
density wave. \hfill} }\label{fig03}
\end{figure}

At the first stage, smooth spiral shocks are formed
on one of the slopes of the potential well: inside the
corotation radius (Fig. 3), these shocks are located
at the leading (with respect to the onflowing gas)
edge of the spiral potential well (line \textit{S01} in Fig. 3c)
produced by stellar density waves; outside the corotation
radius, the shocks are at the trailing edge of the
potential well (line \textit{S02}).

\begin{figure}[!t]
 \setcaptionmargin{5mm} \onelinecaptionstrue
\begin{tabular}{cc}
  \includegraphics[width=0.5\hsize]{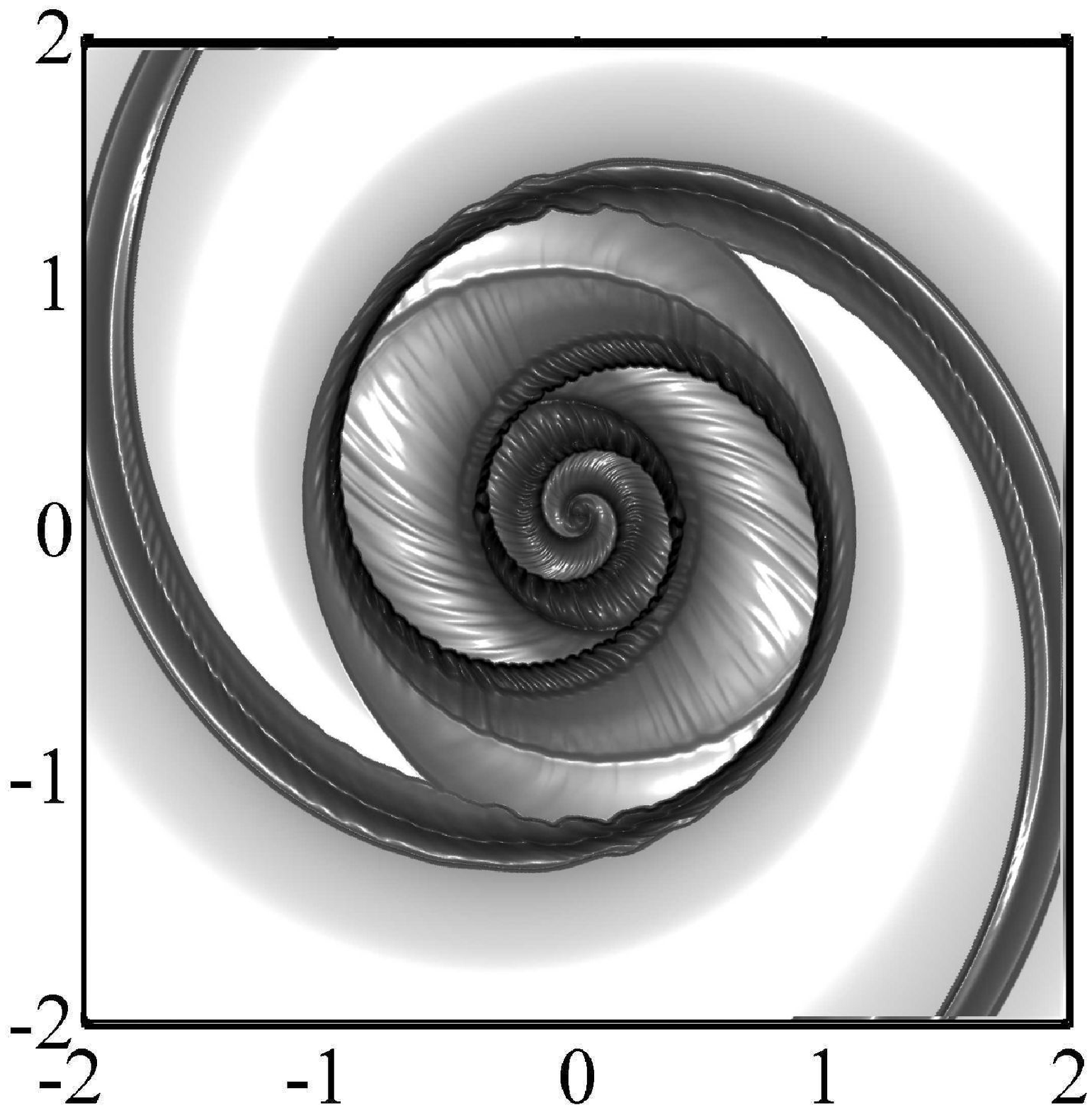} & \includegraphics[width=0.5\hsize]{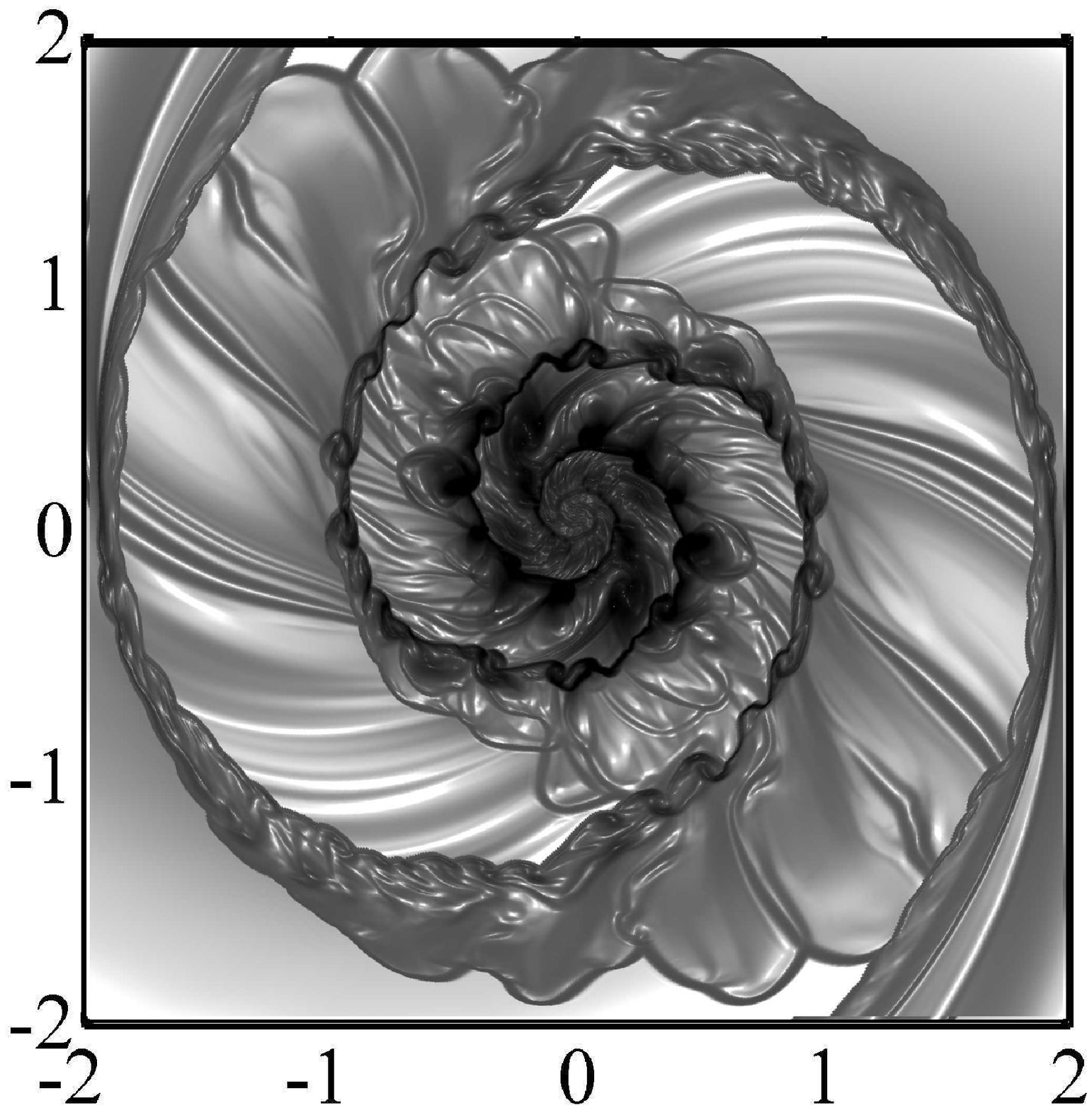}
\end{tabular}
\captionstyle{normal}
 \vskip -0.\hsize \vbox{\hsize=0.95\hsize  \caption{
The model with $\varepsilon_0=0.15$, ${\cal M}_0=30$, $i=15^\circ$, $r_c=4$. (a) The initial stage of wiggle instability development. (b) The stage of developed spurs. A special method is used for visualization: the surface of the function $z=\sigma(x,y)$is shown in the presence of illumination, which enhances the contrast.
\hfill} }\label{fig04}
\end{figure}

At the second stage, small-scale perturbations
giving rise to spurs develop fairly rapidly due to the
wiggle instability of shocks (Fig. 4). The small-scale irregular features in Fig. 2 are clearly traceable starting
from $t=1.5$. This stage can be absent at certain
parameters. = 1.5. This stage can be absent at certain
parameters. The density nodes are clearly seen in
Fig. 4b; the nonlinear wave and the ⌠fish fins■ moving
off the spiral wave at an angle close to 90? consist of
their sequence. The formation of spurs in the spiral
pattern (the galaxies M\,51 and M\,81 are classical
examples) through the development of a nonlinear
wiggle instability stage was considered by Wada and
Koda (2001) and Wada (2008). The formation of a
small-scale structure of spirals turns out to be also
possible in MHD models (Shetty and Ostriker 2006) through the development of gravitational instability
(Shetty and Ostriker 2008) when the thermal effects
are taken into account (Dobbs and Bonnell 2006; Kim
et al. 2008).

\begin{figure}[!t]
 \setcaptionmargin{5mm} \onelinecaptionstrue
\begin{tabular}{ ccc }
  \includegraphics[width=0.34\hsize]{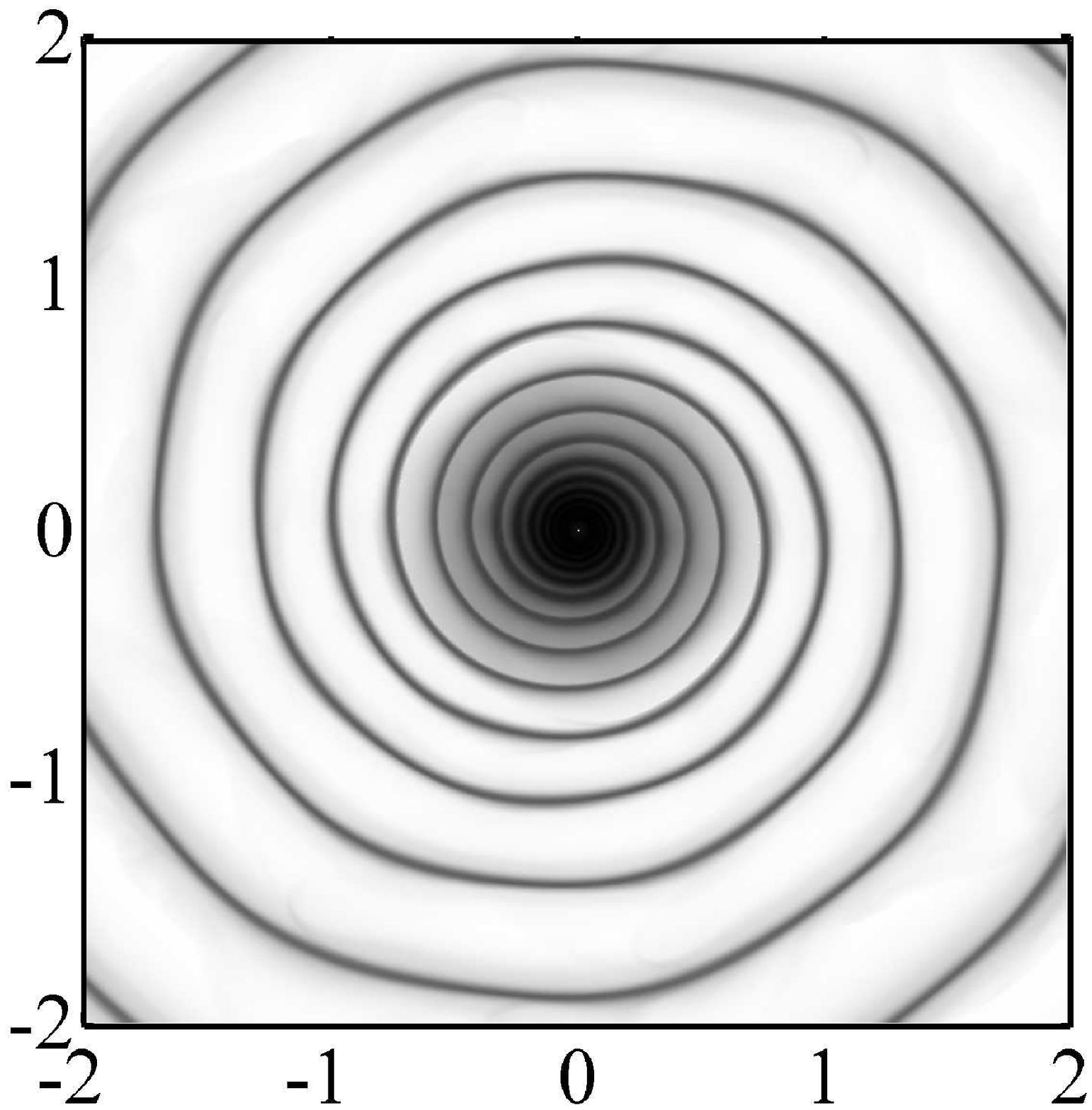} &
  \includegraphics[width=0.34\hsize]{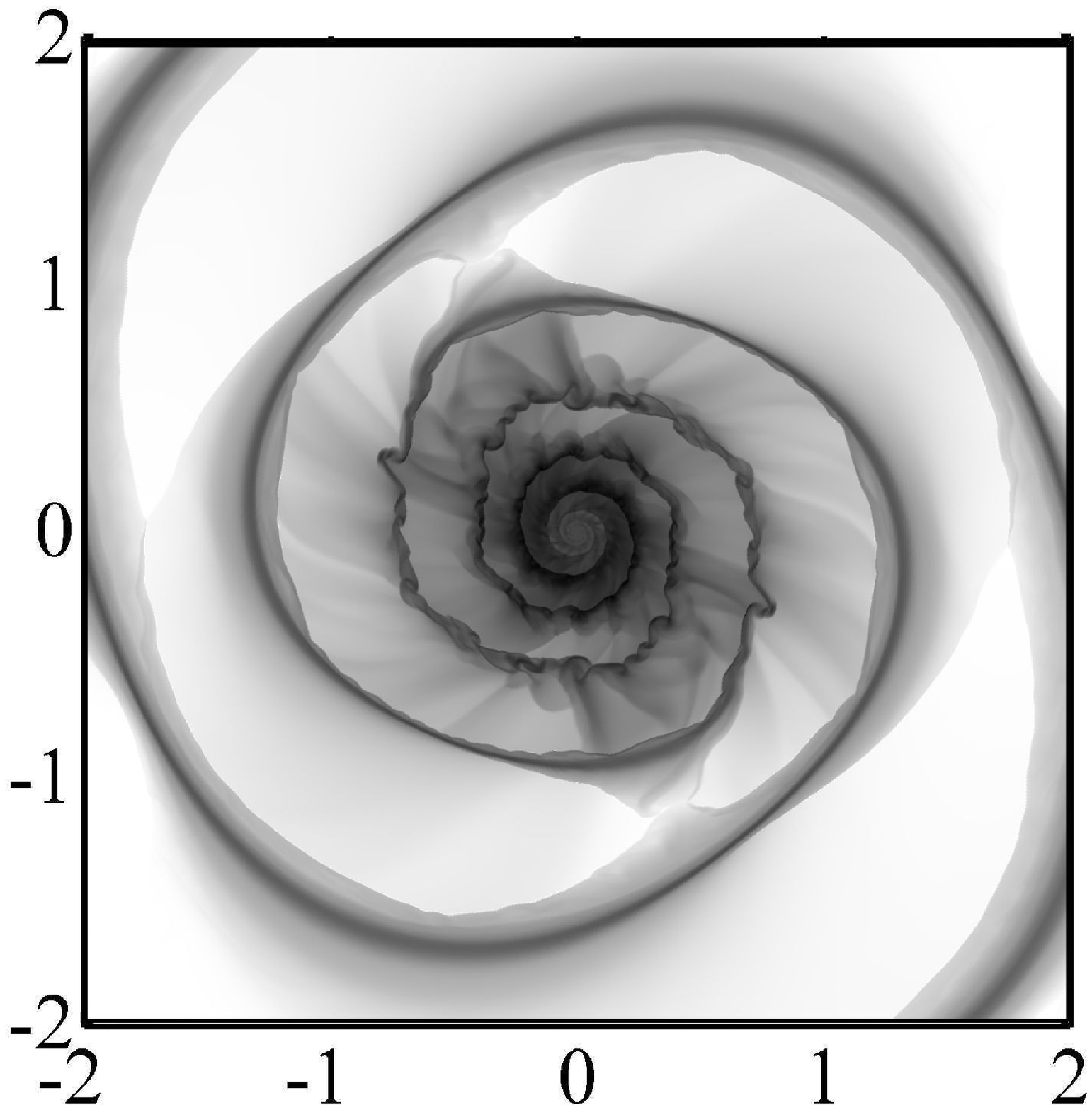} &
  \includegraphics[width=0.34\hsize]{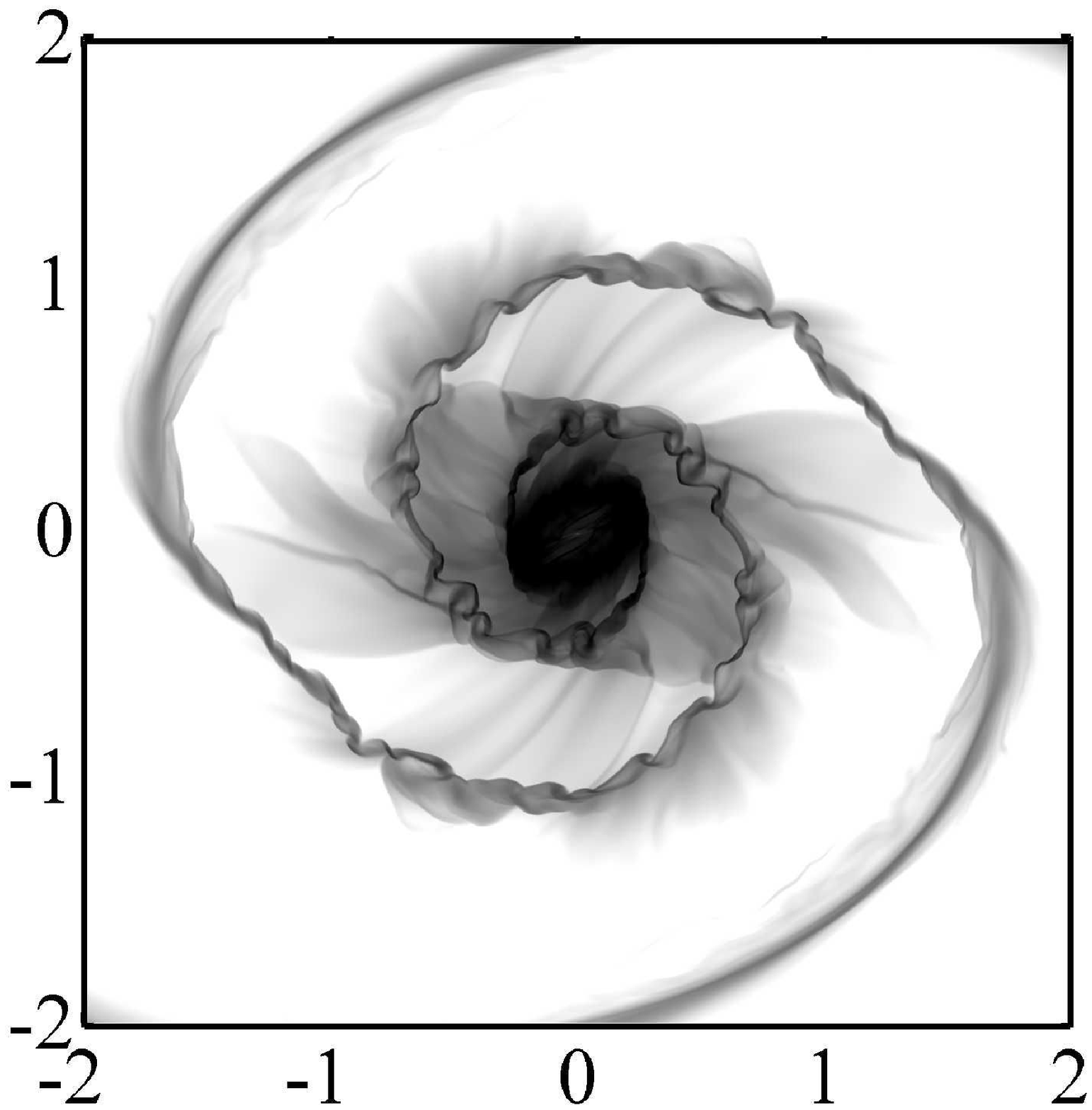} \\
  \includegraphics[width=0.34\hsize]{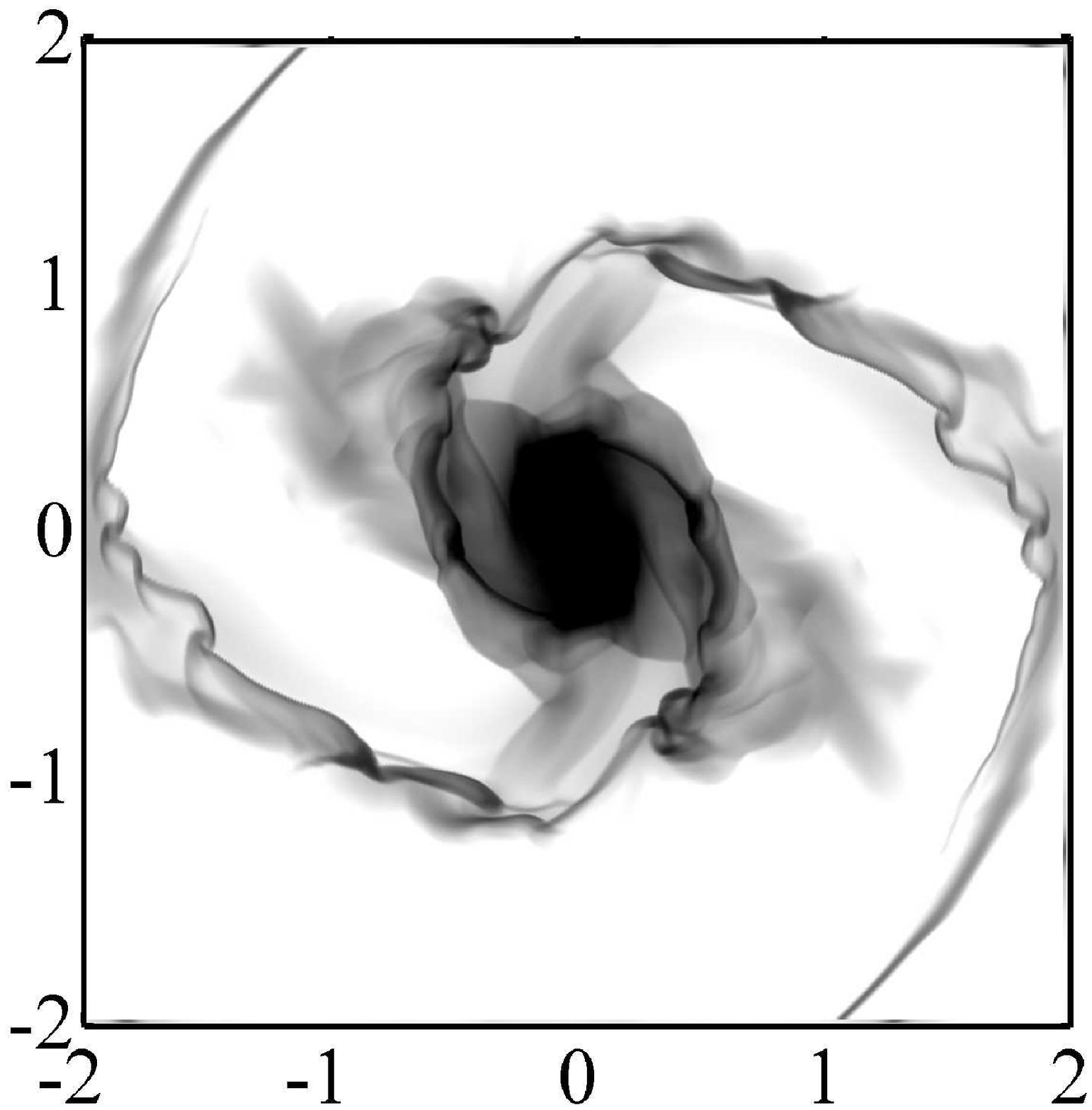} &
  \includegraphics[width=0.34\hsize]{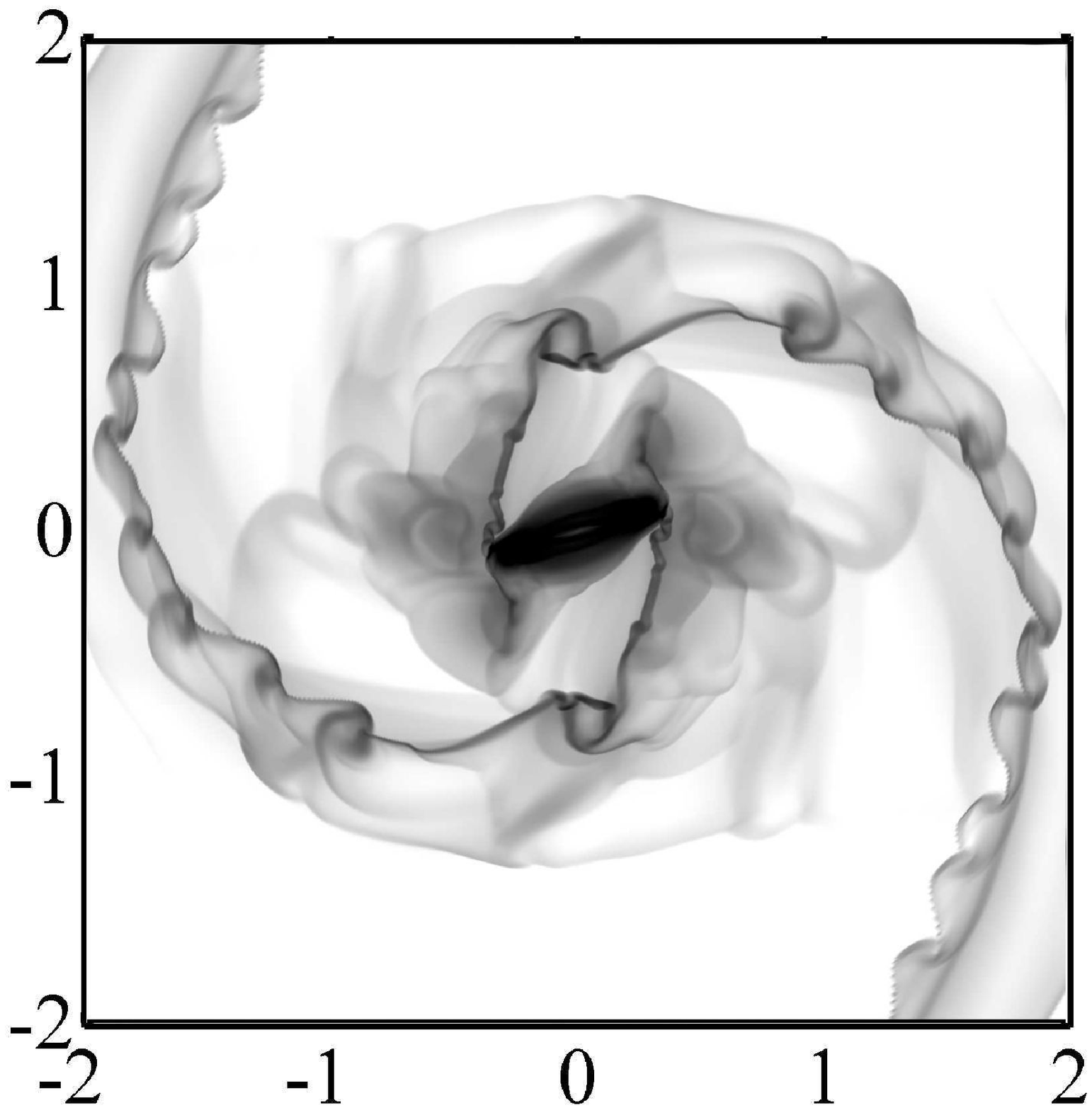} &
  \includegraphics[width=0.34\hsize]{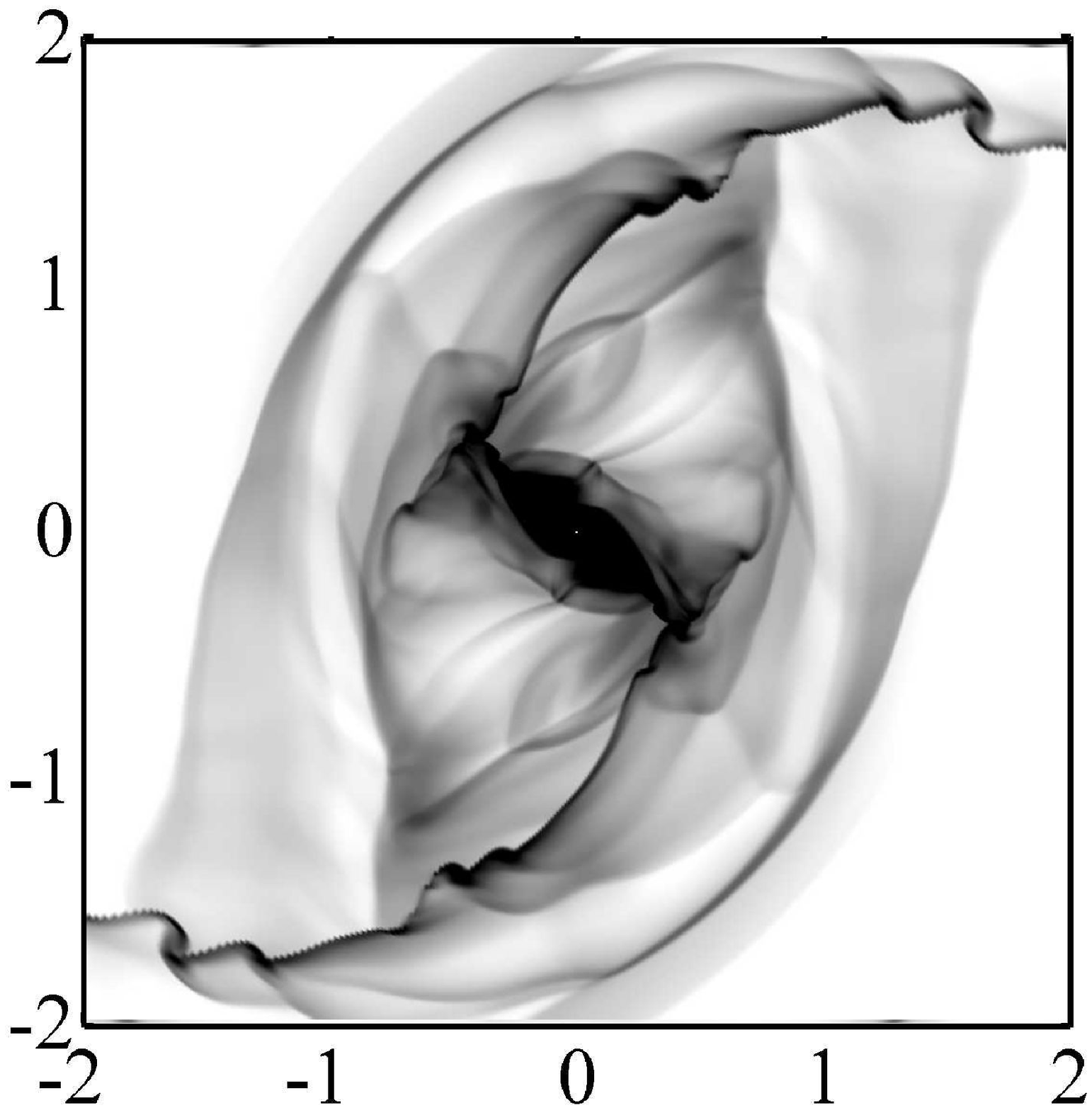} \\
\end{tabular}
\captionstyle{normal}
 \vskip -0.\hsize \vbox{\hsize=0.9\hsize  \caption{
PSs in the models with $\varepsilon_0=0.1$, ${\cal M}_0=30$, $r_c=4$ for various pitch angles $i= 5^\circ; 10^\circ; 15^\circ$(lower row).
\hfill} }\label{fig05}
\end{figure}

At the third stage, the effect of spiral shock escape
from the potential well is observed and the straightening
of shock segments giving rise to polygonal
structures (rows) in the gaseous disk emerges under
certain conditions to be discussed below (see Figs. 2,
3c, 5). Such straight fragments generally follow the
geometry of the stellar spiral density wave that is determined
by the nonaxisymmetric part of potential (\ref{Pot}). The evolution of the spiral structure in gas is peculiar
in that it is nonstationary≈the rows do not form any
stationary pattern, which is additionally complicated
by the system of spurs whose properties change on
short time scales compared to the disk revolution
period.

The shock escape from the potential well in the
constructed models turns out to be possible only if
the shock is on the outer slope of the potential well
(see \textit{S01} in Fig. 3c)≈the transition to position \textit{SP1} occurs. The transition of shock \textit{SP2} from the inner slope of the potential well to the polygonal geometry
SP2 turns out to be impossible. The former situation
(\textit{S01}$\rightarrow$\textit{SP1}) corresponds to the outer corotation radius $r_{c1}$  (see Fig. 3b), while the latter situation corresponds to fast rotation of the spiral pattern with
$r_{c2}$. Thus, in the purely hydrodynamic model being
discussed here, the position of the corotation radius
on the disk periphery is a necessary condition for the
emergence of PSs.

The described result differs from the conclusion
reached in Chernin et al. (2006), where case \textit{SP2} is realized (see Fig. 3c). The consideration of only
a plane flow in Chernin et al. (2006) without any
allowance for disk rotation, radial inhomogeneities of
the spiral potential and gas density, and rotation difis realized (see Fig. 3c). The consideration of only
a plane flow in Chernin et al. (2006) without any
allowance for disk rotation, radial inhomogeneities of
the spiral potential and gas density, and rotation differentiality
is responsible for this difference. Including
the latter factors changes radically the position of the
shock front in the potential well. In the plane problem,
the shock is at the entrance of a supersonic flow into the potential well. In the realistic model of a rotating
gaseous disk and a spiral stellar density wave, the
shock is at the exit from the potential well.

We constructed a series of models with various
spiral pitch angles $i$ (see Fig.~5). At any values
of $i$ in our numerical simulations, we can identify the
time intervals with well-defined rows even for a small
amplitude of the stellar density wave, $\varepsilon_0
 =0.1$. At small pitch angles (see the case ofz$i= 5^\circ$ in Fig.~5), the formation of PSs is not accompanied by the development
of wiggle instability.

\begin{figure}[!t]
 \setcaptionmargin{5mm} \onelinecaptionstrue
\begin{tabular}{ccc}
  \includegraphics[width=0.34\hsize]{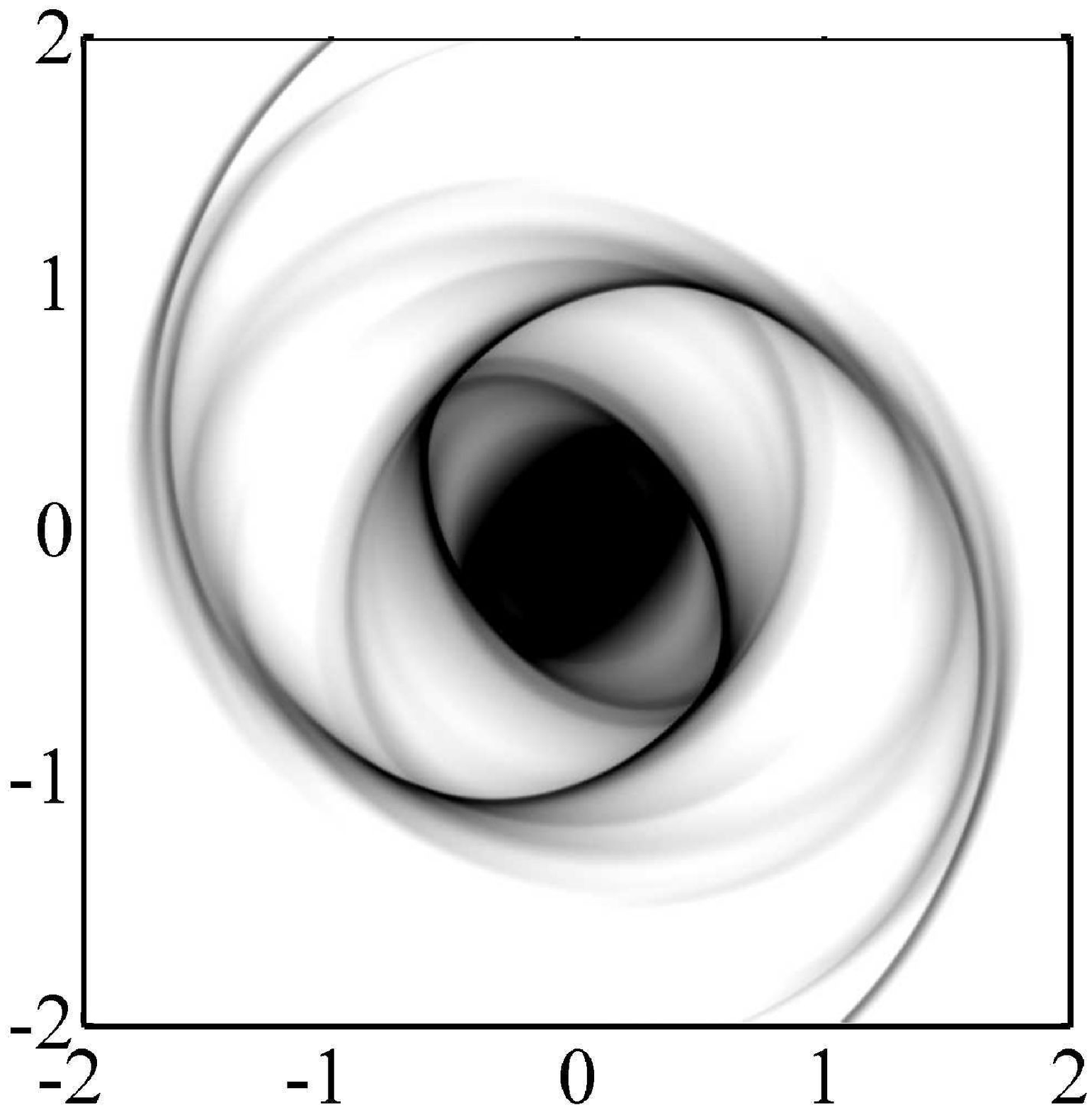} &
  \includegraphics[width=0.34\hsize]{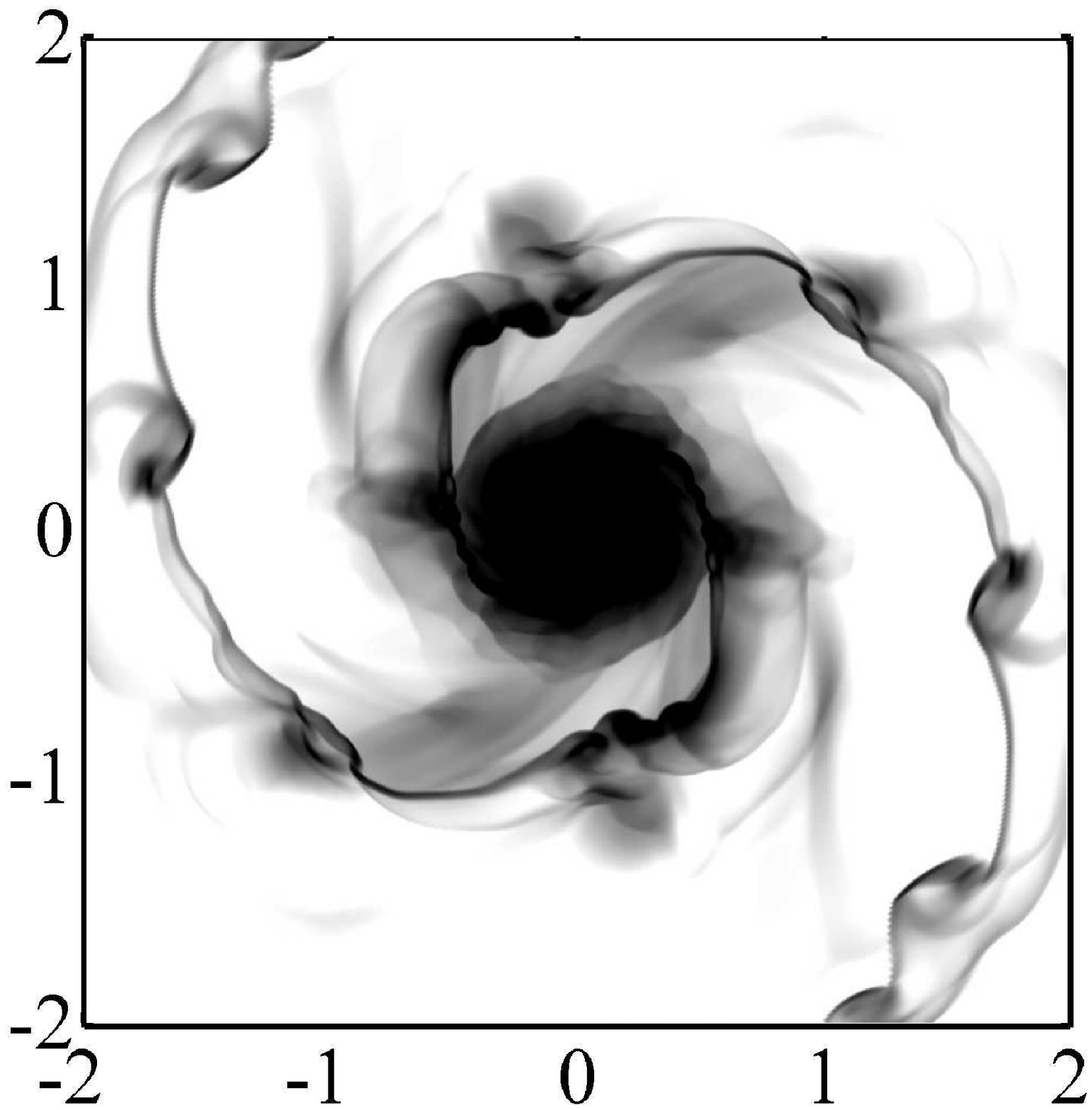} &
  \includegraphics[width=0.34\hsize]{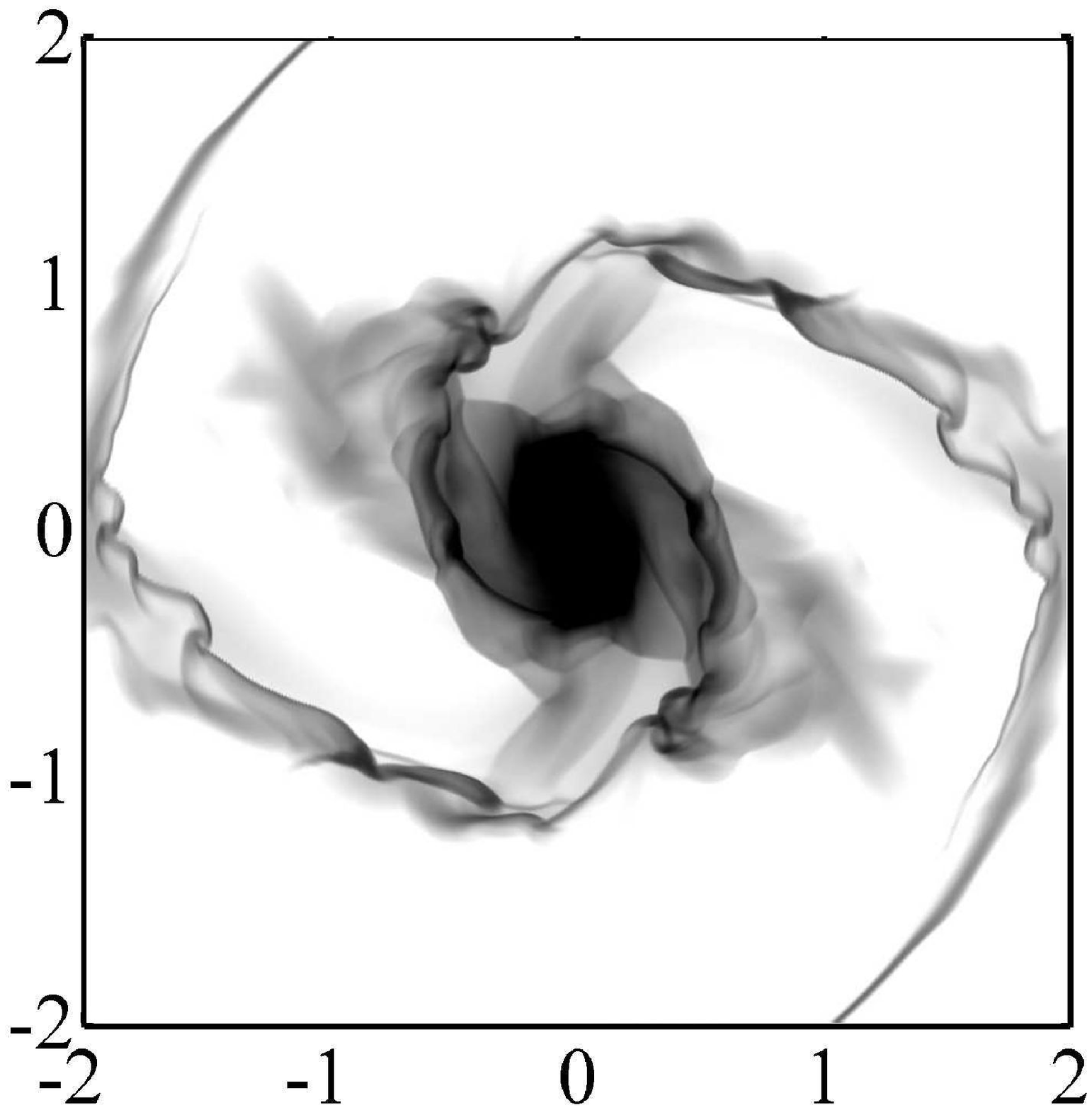} \\
  \includegraphics[width=0.34\hsize]{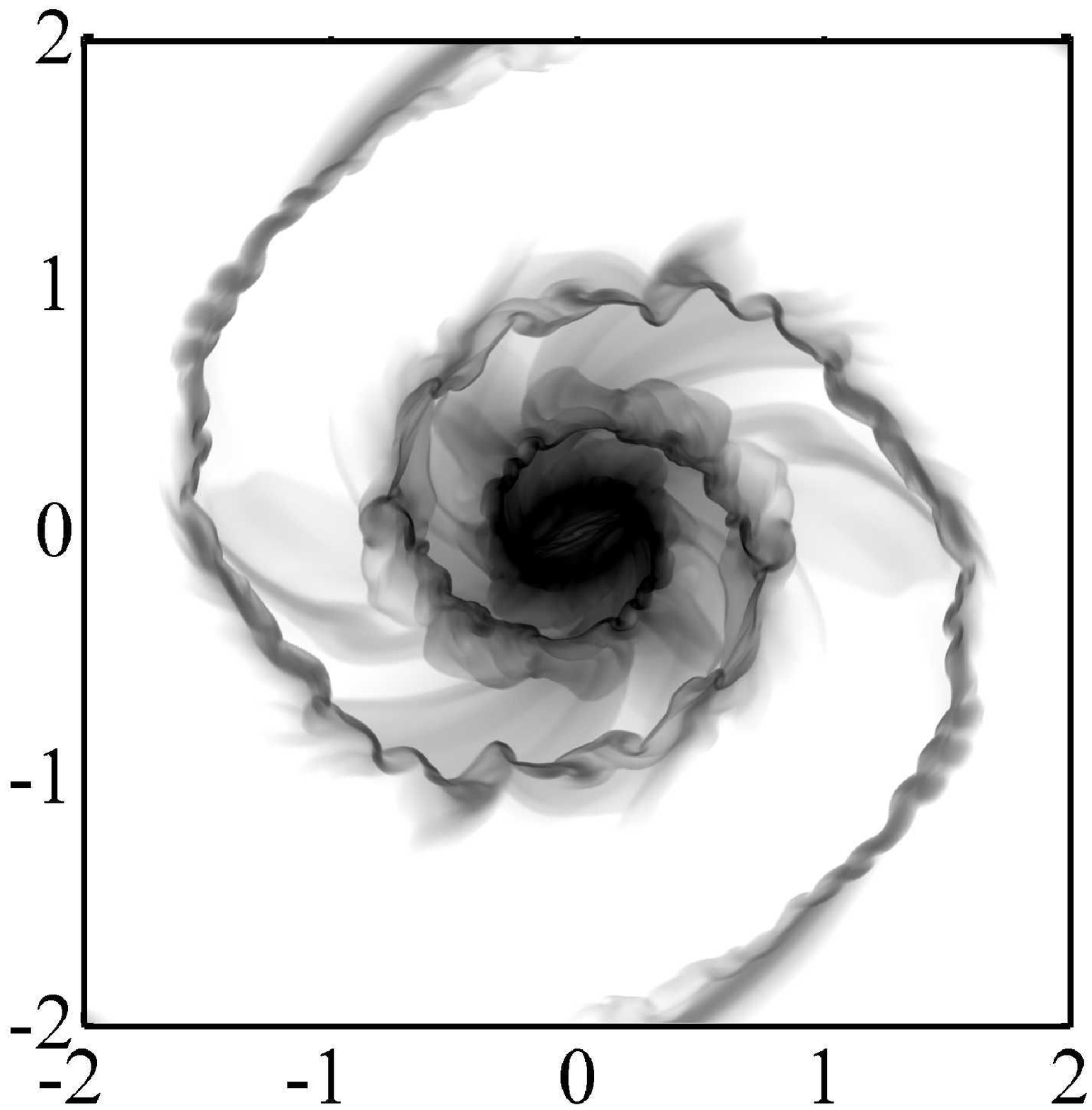} &
  \includegraphics[width=0.34\hsize]{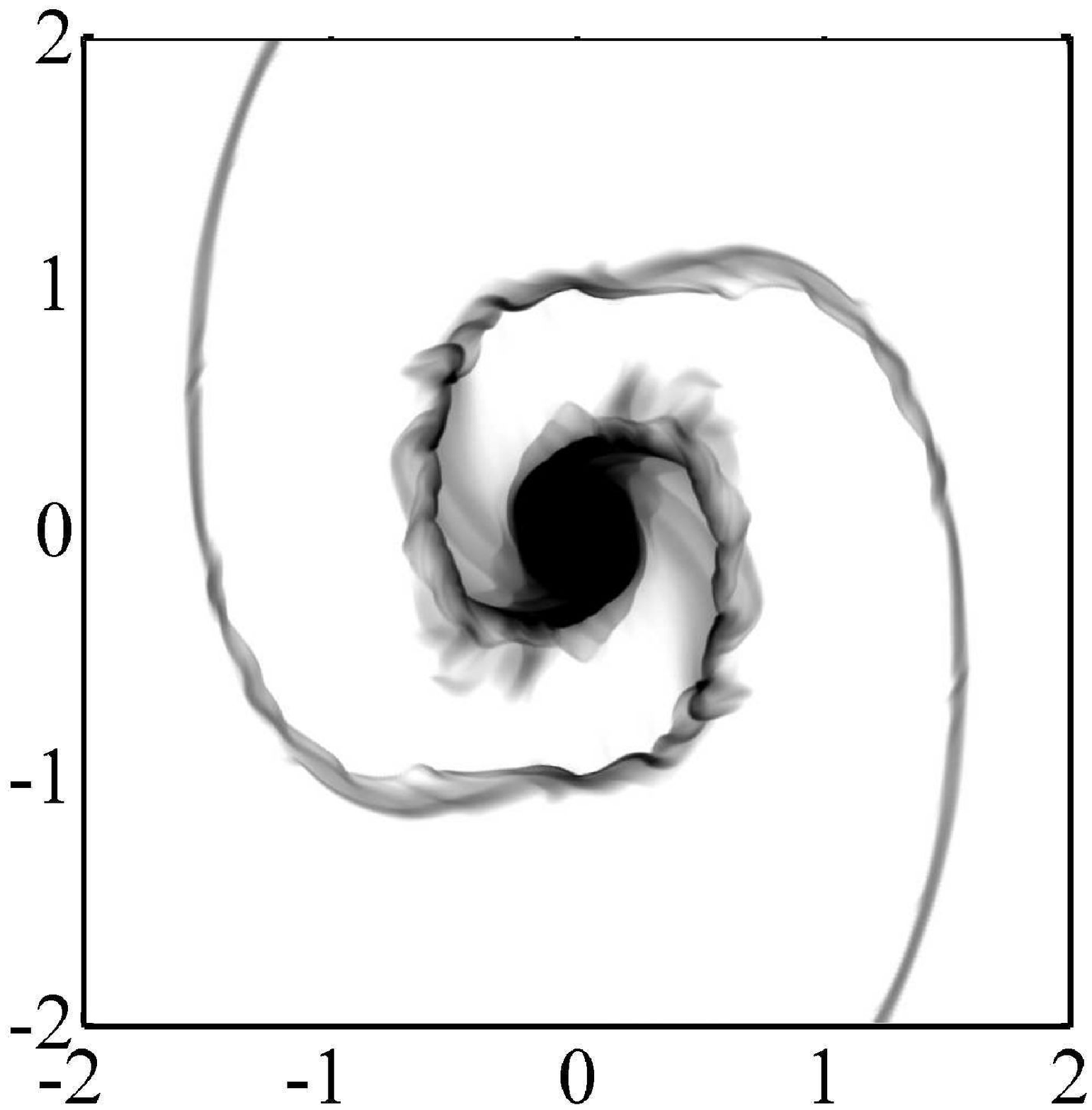} &
  \includegraphics[width=0.34\hsize]{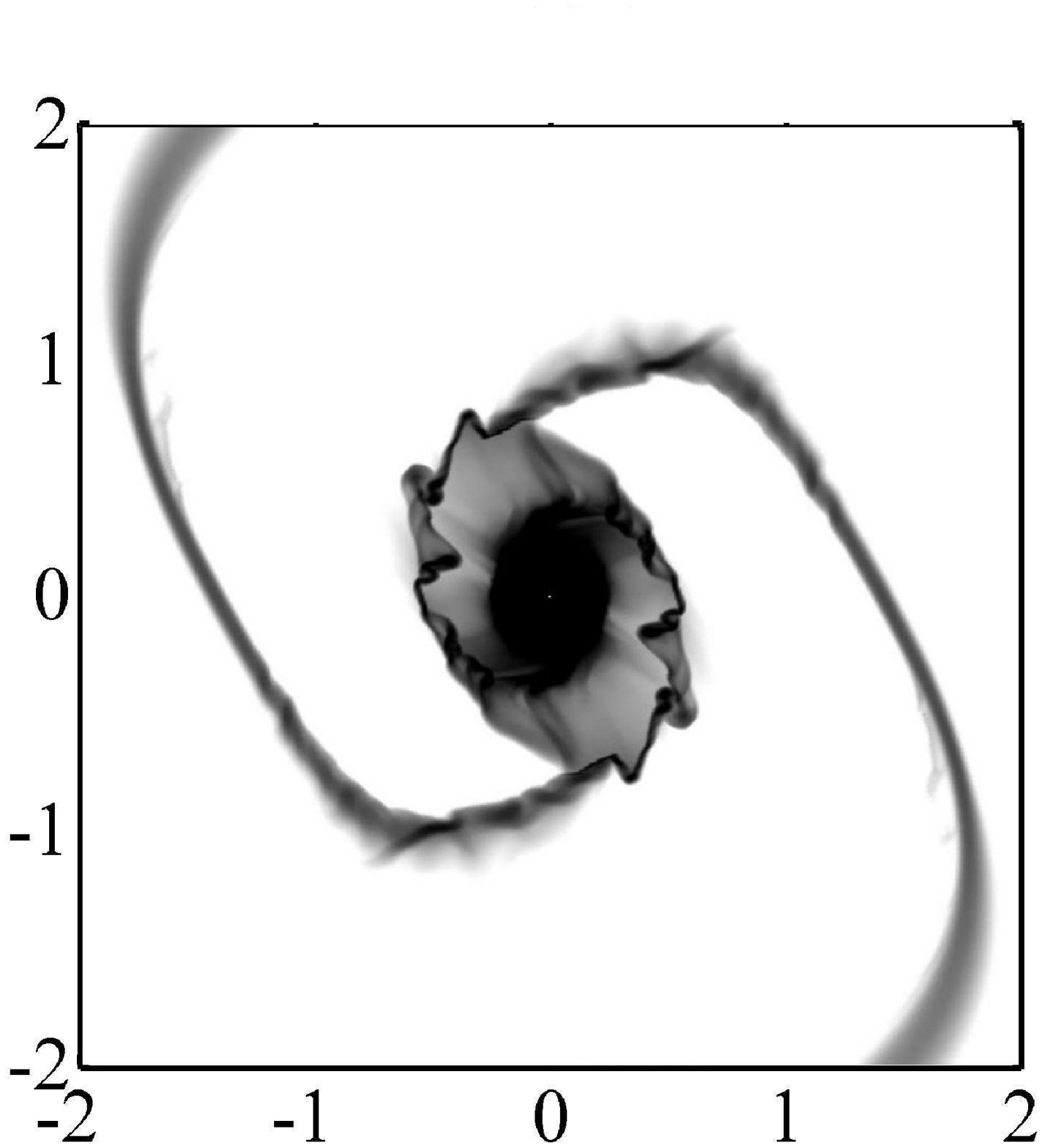} \\
\end{tabular}
\captionstyle{normal}
 \vskip -0.\hsize \vbox{\hsize=0.9\hsize  \caption{The models with  $i= 20^\circ$, ${\cal M}_0=30$, $r_c=4$ for various values of $\varepsilon_0=0.01; \ 0.05; \ 0.1$ (upper row); $\varepsilon_0=0.15; \ 0.2; \ 0.3$ (lower row).
\hfill} }\label{fig06}
\end{figure}

The amplitude of the stellar spiral density wave $\varepsilon_0$
affects significantly the formation of the spiral structure
in the gas subsystem (Fig.~6). The amplitude of
the stellar spiral density wave affects in a complex way
the PS generation efficiency. On the one hand, the
wave amplitude in gas increases with $\varepsilon_0$. On the other
hand, it is easier for the shock to escape from a shallow
shock potential well, which is necessary for the
formation of straightened segments of perturbations.
As we see, the former factor plays a more important
role for the formation of PSs. In a deeper potential
well (at larger $\varepsilon_0$), we obtain more pronounced kinks
of the shock front in the gas component. At a small
amplitude  $\varepsilon_0\lee 0.01$, no rows are formed in our simulations (see Fig.\,6).

The transient nature of PSs is a characteristic
feature of the constructed numerical models. Their
parameters (the positions of kinks, the lengths of
rows, the angle between them, and even their number)
change with time. Having emerged after the
shock escape from the potential well, the PSs begin
to weaken, almost disappearing, and the rows are
subsequently restored again. In the case of a deeper
potential well, the rows live over longer time intervals.
The generation of PSs is determined by the physics
of shocks in a spiral gravitational well, which begin to
evolve due to the development of shear instability.

The constructed models are not self-consistent,
because no allowance is made for the reverse effect
of the gas on the geometry of the spiral pattern in the
stellar component. In the case of such an allowance,
one would expect an enhancement of the transient
character of PSs.

\begin{figure}[!t]
 \setcaptionmargin{5mm} \onelinecaptionstrue
\includegraphics[width=0.85\hsize]{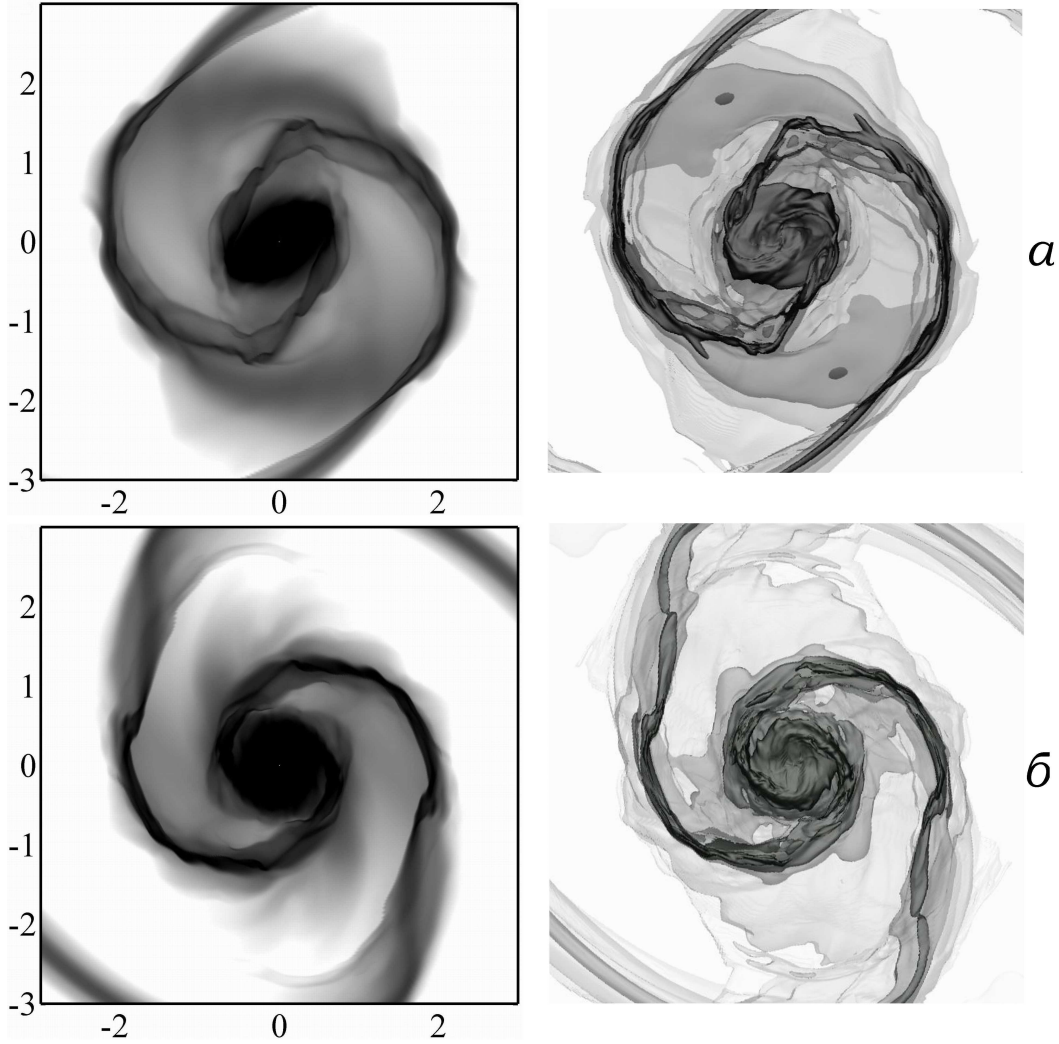}
\captionstyle{normal}
 \vskip -0.\hsize \vbox{\hsize=0.9\hsize  \caption{
The structure of a 3D gaseous disk in the presence of rows at two different instants of time (a, b): the surface density is
on the left; the isosurfaces of the logarithm of volume density are on the right.
\hfill} }\label{fig07}
\end{figure}

 On the whole, our 3D simulations in the cylindrical
coordinate system on a grid with a resolution
 $\Delta{r}\times \Delta\varphi \times
\Delta{z}= 0.01\times1^\circ\times0.01$ confirm the
results of our 2D models (Fig.~7). Transient PSs are
formed in the same way.

%----------------------- Section 3 -------------------------------

\section{PARAMETERS OF POLYGONAL
STRUCTURES}

\noindent

The formation of straightened segments of spiral
shocks depends on model parameters. Let us discuss
the pattern of this influence.

First, the number of PSs depends on the spiral
pitch angle. At small pitch angles, $i=5^\circ -20^\circ$, the
number of straightened segments is great (typically
five√seven or more). At large pitch angles, $i\geq50^\circ$,
the number of kinks decreases to one or two.

Second, slow rotation of the spiral pattern at which
the corotation radius is on the disk periphery (see
Fig.~3) contributes best to the formation of PSs.
When $r_c=V(r_c)/\Omega_p \lesssim L_\sigma$, no rows are observed.

Third, the evolution of the gaseous disk is strongly
affected by the depth of the spiral potential well $\varepsilon_0$.
At low perturbation amplitudes, $\varepsilon_0\lesssim 0.03$, no wiggle
instability develops; nevertheless, the straightening
of the spiral shock front turns out to be possible. In
models with large depths of the potential well, $\varepsilon\geq0.25$, almost stationary PSs are formed in the gaseous disk. In this case, not only the shock escape from the
potential well but also the spiral shock splitting into
two (or occasionally more) branches is observed.

\begin{figure}[!t]
 \setcaptionmargin{5mm} \onelinecaptionstrue
\begin{tabular}{ c c }
  \includegraphics[width=0.55\hsize]{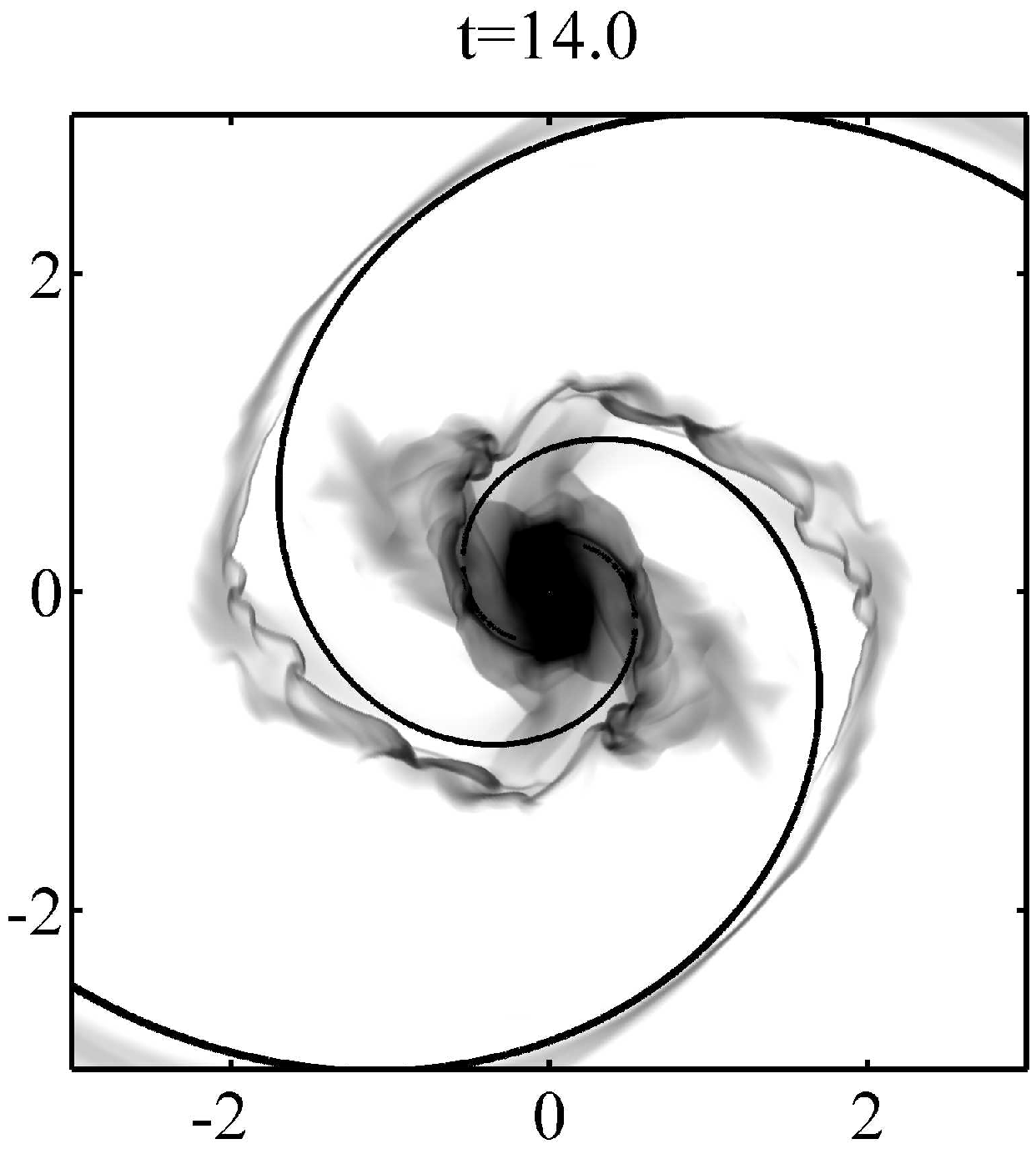} &
  \includegraphics[width=0.55\hsize]{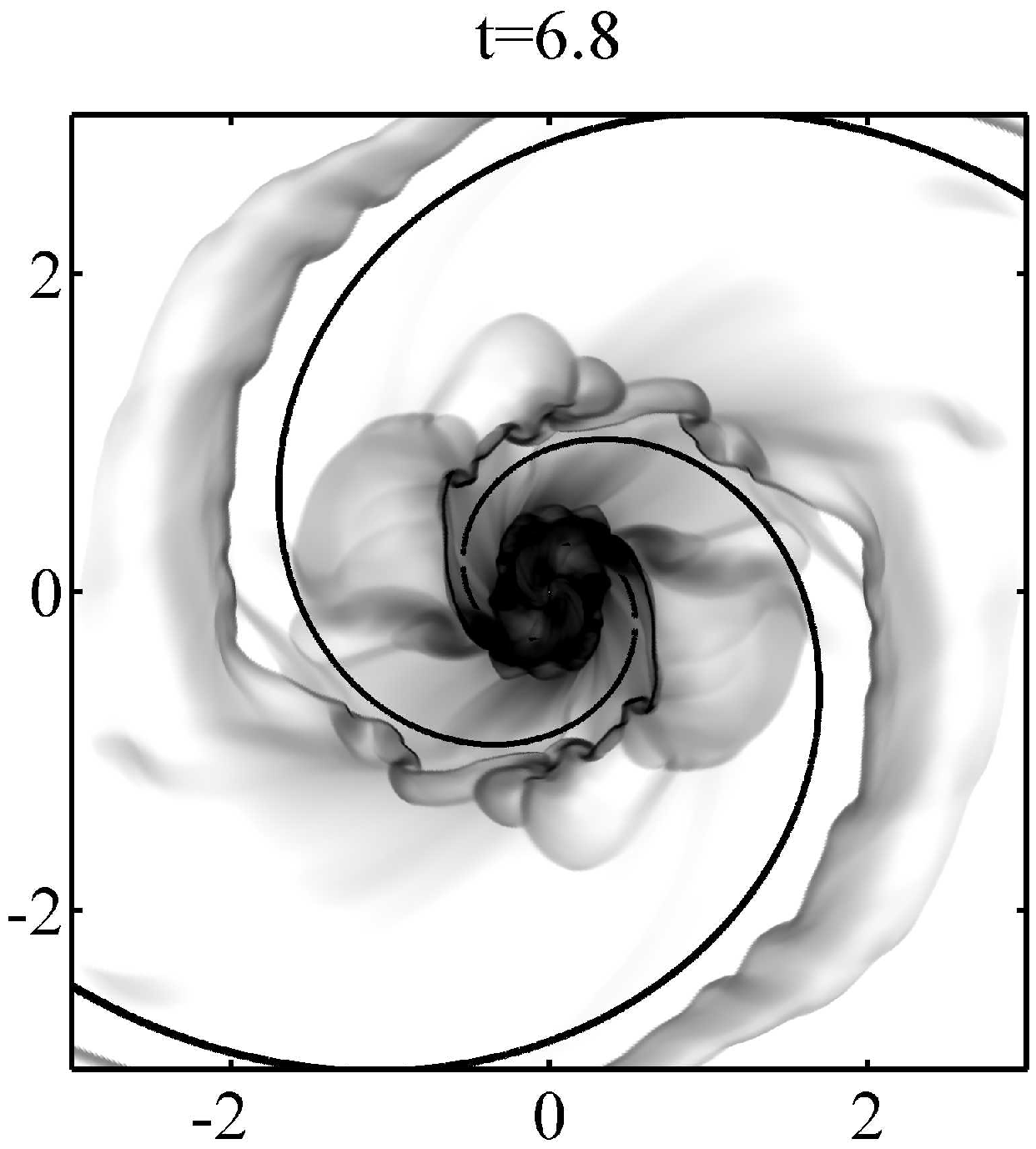}
\end{tabular}
\captionstyle{normal}
 \vskip -0.\hsize \vbox{\hsize=0.9\hsize  \caption{
PS positions relative to the minimum of the potential well fromthe spiral arm of the stellar component (solid lines). The
models with ${\cal M}_0=30$, $i=20^\circ$, $\varepsilon_0=0.1$, $\Omega_p=0.5$ (a), $\Omega_p=0.1$ (b).
\hfill} }\label{fig08}
\end{figure}

In the models considered, the rows are always
located at the outer edge of the potential well≈ahead
(with respect to the gravitational potential minimum)
of the stellar density wave, provided that the corotation
radius $r_c$ lies on the disk periphery. In this
case, increasing the corotation radius from $r_c=4$
(Fig.~8a) to $r_c=20$ (Fig.~8b) retains the possibility
of the formation of rows.

Note the influence of the Mach number ${\cal M}_0$ on
PS properties. All other things being equal, fast disk rotation (large ${\cal M}_0$) contributes to the formation
of PSs. At effective Mach numbers ${\cal M}_0\lesssim 10$, the
conditions for the development of wiggle instability
deteriorate. This, in turn, makes the shock escape
from the gravitational well of a stellar density wave
difficult, which is necessary for the formation of an
extended straightened segment of the gaseous pattern.
If we take $c_s\sim 10-15$ $\text{km s}^{-1}$, then we have
$V_{\max}\simeq 100 - 150$ $\text{km s}^{-1}$ for ${\cal M}_0=10$. Note that
rows are encountered very rarely among slowly rotating
galaxies.

Figure 9 shows the PS parameters from more
than 40 numerical simulations in comparison with
the observational data for 200 galaxies (Chernin et al.
2001a). For each simulation, we chose several instants
of time with a developed system of rows. The
total number of measurements was 450. The histograms
of the number of straightened spiral arm
segments $N_g$ versus angle between neighboring segments
$\alpha$ are shown at the top. As we see, the maximum
in the distribution occurs at an angle of $120^\circ$, in
complete agreement with the observational data. The
number of pairs of rows with $\alpha=(120\pm 2.5)^\circ$ in our
numerical simulations was 30\% of the total number.

\begin{figure}[!t]
 \setcaptionmargin{5mm} \onelinecaptionstrue
\begin{center}
\includegraphics[width=0.45\textwidth, keepaspectratio]{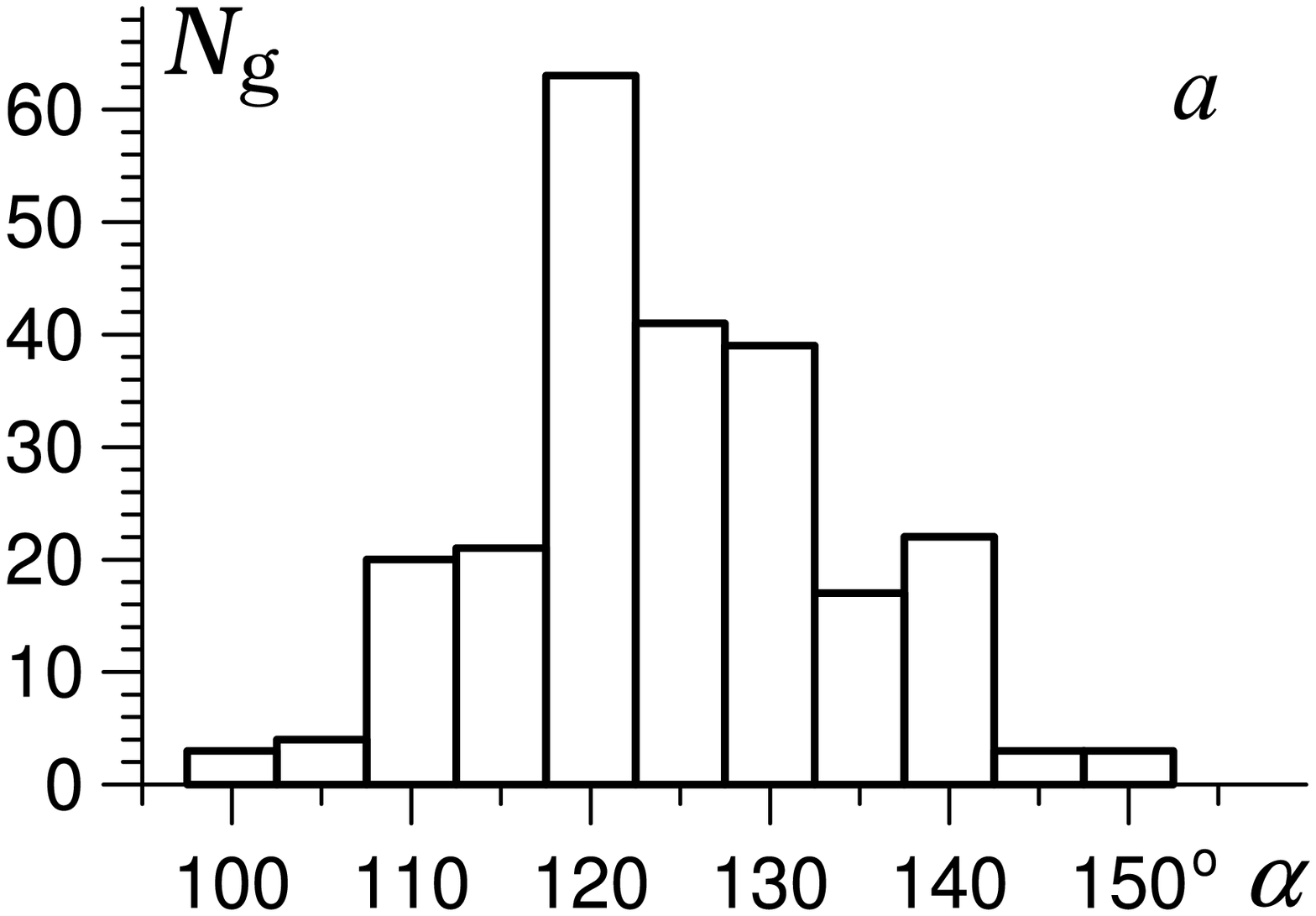}
\hskip 3 true mm
\includegraphics[width=0.45\textwidth, keepaspectratio]{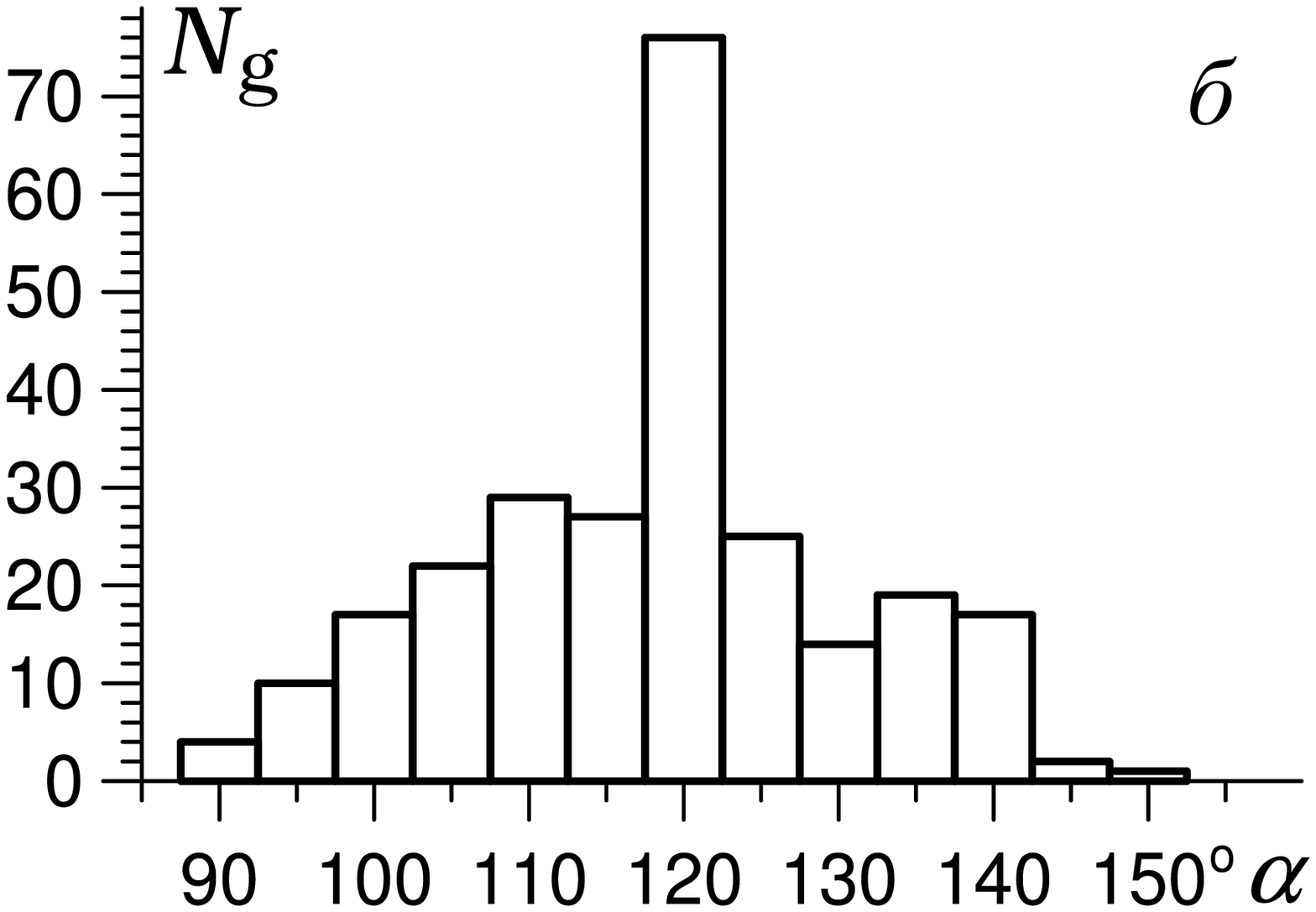}
\\
\includegraphics[width=0.45\textwidth, keepaspectratio]{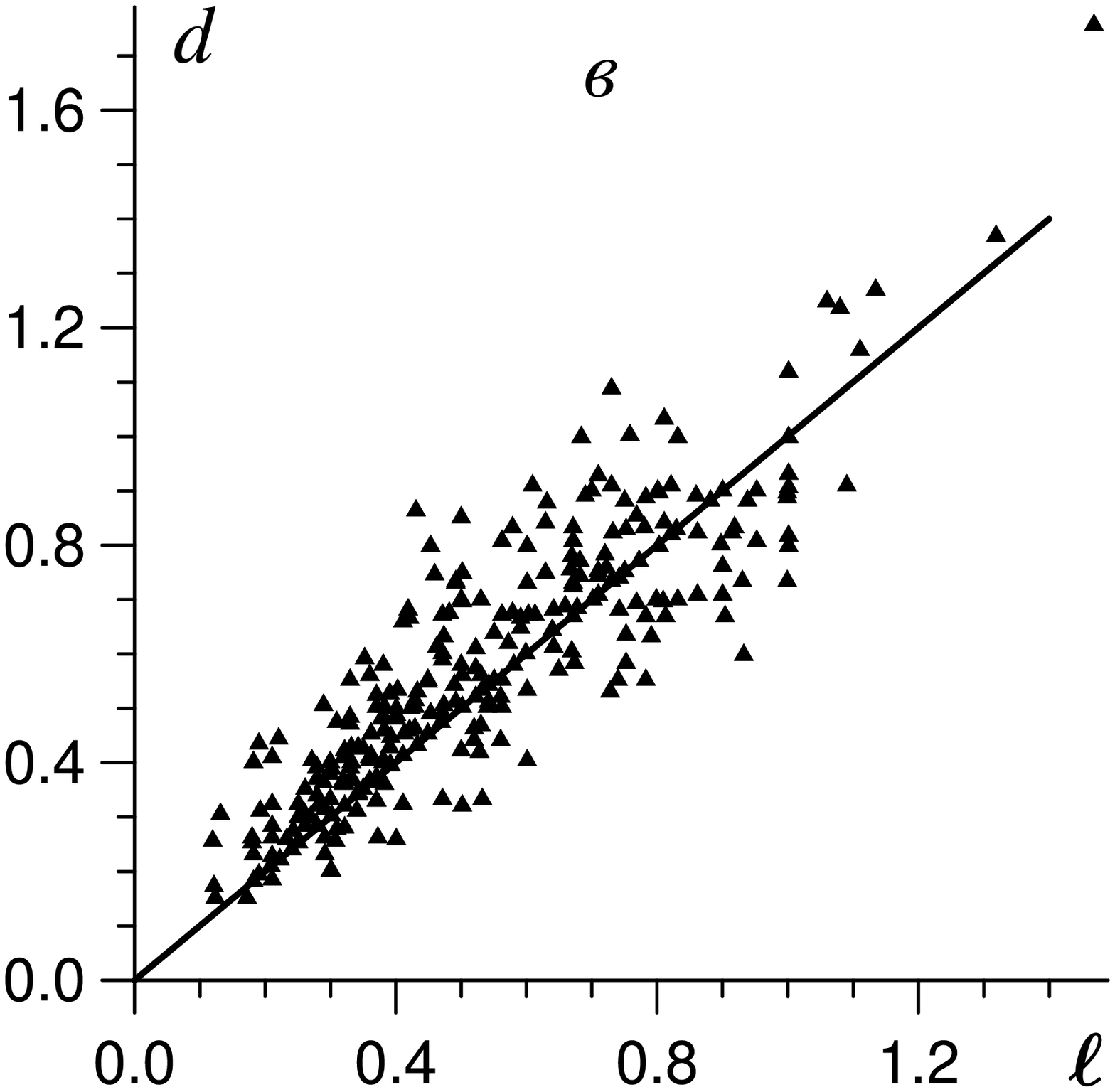}
\hskip 3 true mm
\includegraphics[width=0.45\textwidth, keepaspectratio]{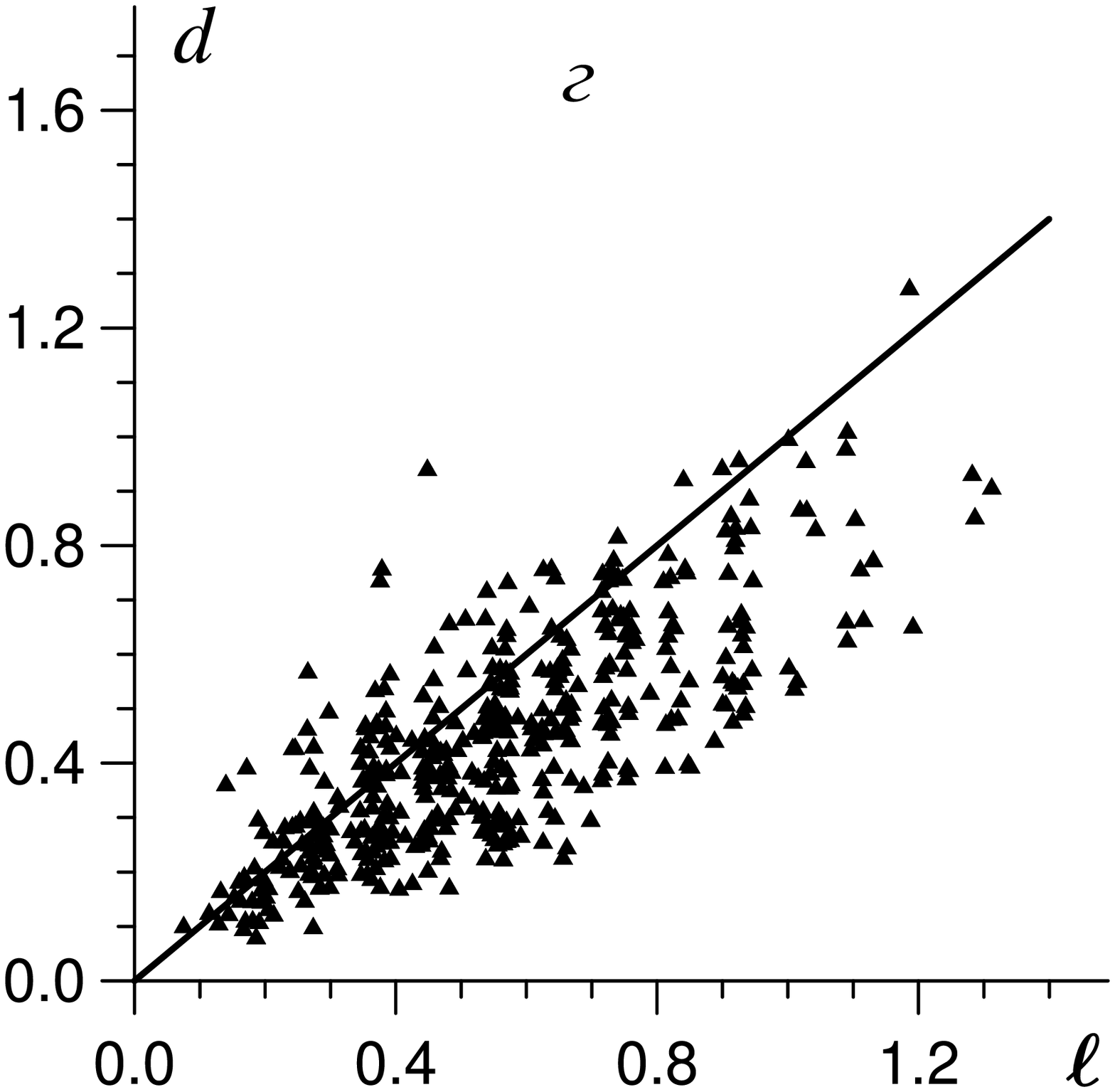}
\\
\end{center}
\captionstyle{normal}
 \vskip -0.\hsize \vbox{\hsize=0.95\hsize  \caption{
Results of the statistical analysis of observational data for a sample of stars (Chernin et al. 2001a) (a, c) and for a series
of numerical simulations at various instants of time at the stage with a developed system of rows (b, d): (a, b) Histograms of
the number of straightened segments versus angle $\alpha$ between them; (c, d) length $\ell$ of a straightened segment versus distance $d$ to the disk center. The solid line: $d=a\ell+b$ obtained by Chernin et al. (2001a). \hfill} }\label{fig09}
\end{figure}

The size of a straightened spiral segment depends
on Galactocentric distance linearly, but the slope of
the straight line turns out to be slightly different from
the value obtained by processing the observational
data in Chernin et al. (2000). This difference may
stem from the fact that we used a cube of models
with a uniform step in parameters $i,\varepsilon_0,{\cal
 M}_0$, while the sample of galaxies definitely does not cover uniformly the grid of parameters ($i,\varepsilon_0,{\cal
 M}_0$). On the other hand, we did not consider spiral patterns withz $m>2$ arms, while a significant fraction of the total number of objects with rows in the sample of galaxies is accounted
for by multiarmed patterns.

Significantly, the PS formation mechanism considered
here is purely hydrodynamic and is associated
with an unstable location of the shock front in the
spiral potential well.

%йЮЙ НРЛЕВЮКНЯЭ БШЬЕ, Х ДПСЦХЕ ЛЕУЮМХГЛШ ЛНЦСР
%БКХЪРЭ МЮ ДХМЮЛХЙС СДЮПМШУ БНКМ.

%----------------------- Section 3 -------------------------------

\section{CONCLUSIONS}

\noindent

In conclusion, let us formulate our main results:
\begin{itemize}
 \item[1.] We showed that PSs could be formed in the gas
component in a simple hydrodynamic model due to an
unstable position of the shock front in a spiral stellar
density wave. Numerical models with high spatial
and temporal resolutions are needed to generate these
features.

 \item[2.] We showed that PSs could be formed for pitch
angles of the spiral pattern $5^\circ - 50^\circ$.
 \item[3.] As a rule, the system of rows in the numerical
models is essentially nonstationary.
 \item[4.] The characteristic formation time of straightened
segments does not exceed the revolution period
of the stellar disk periphery.
 \item[5.] PSs emerge in models with a slowly rotating
spiral pattern when the corotation radius is on the
disk periphery.
 \item[6.] The results of our 2D and 3D gaseous disk
simulations are in good agreement.
  \item[7.] The geometric properties of rows in the models
agree well with the processing results for observed
spiral patterns. In particular, the predominant angle
between segments is ${\alpha}\simeq 120^\circ$.
  \end{itemize}

The increase in the fraction of galaxies with PSs
among the interacting ones probably stems from the
fact that an external gravitational effect facilitates the
escape of a galactic shock from the potential well of a
stellar arm, facilitating the shock straightening.

\section*{ACKNOWLEDGMENTS}
We wish to thank A.V. Zasov, V.I. Korchagin,
and A.D. Chernin for useful discussions. The numerical
simulations were performed on the ⌠Chebyshev■
supercomputer of the Moscow State University
with assistance from A.V. Zasov and N.V. Tyurina.
This work was supported (project no. 10-07-97017-
Povolzh▓e) in part by the ⌠Scientific and Scientific-
Pedagogical Personnel of Innovational Russia■ Federal
Goal-Oriented Program (P1248) and non-profit foundation "Dynasty" of Dmitry Zimin.

\newpage
%%%\vspace{3cm}

%\selectlanguage{english}
%\begin{center}
%\large \bfseries \MakeTextUppercase{%
%Polygonal Structures in the Gaseous Disk: Numerical Simulations }
%\end{center}
%%
%\begin{center}
%\bfseries S.A. Khoperskov, A.V. Khoperskov, M.A. Eremin,  M.A.
%Butenko
%\end{center}
%
%\begin{center}
%\begin{minipage}{\textwidth - 2cm}
%\small In paper were demonstrated results of numerical simulations
%of gaseous disk in spiral potential of stellar density wave. The
%conditions under which the formation of straightened segments of
%the spiral arms (or rows) in the gas component were studied. These
%features of the spiral structure were identified in works of
%Chernin A.D. with co-autors. Gas-dynamic simulations were
%performed for a wide range of model parameters: the pitch angle,
%the amplitude of the stellar spiral density wave, rotational speed
%of the disk, temperature of gas component. We also made a
%comparison of results of 2D- and 3D- simulations. It is shown that
%rows in the numerical experiments are essentially time-dependent
%phenomenon. Statistical analysis of the distribution of geometric
%parameters of spiral patterns with rows of the observed galaxies
%and our hydrodynamic models demonstrates good agreement. In
%particular for numerical experiments and observations of galaxies
%we had the average value of angle between rows equal
%$\langle{\alpha}\rangle\simeq 120^\circ$.
%\end{minipage}
%\end{center}


\centerline{REFERENCES}

\begin{thebibliography}{99}

{

 \bibitem{Acharova-etal-2010!metal-corot-MW}
I. A. Acharova, J. R. D. Lepine, Yu. N. Mishurov,
et al.,Mon. Not. R. Astron. Soc. 402, 1149 (2010).

\bibitem{Buta-Zhang-2009!Corotation-radii}
R. J. Buta and X. Zhang, Astrophys. J. Suppl. Ser.
182, 559 (2009).

 \bibitem{Buta-Combes-1996!Ring}
R. Buta and F. Combes, Fund. Cosmic Phys. 17, 95
(1996).

\bibitem{Buta-Purcell-1998!NGC-3081}
R.Buta andG. B. Purcell,Astron. J. 115, 484 (1998).

\bibitem{Chakrabarti-etal-2003!Branch-Spur-Feather}
S. Chakrabarti, G. Laughlin, and F. H. Shu, Astrophys.
J. 596, 220 (2003).

\bibitem{Chernin1998}
A. D. Chernin, Astrophys. 41, 399 (1998).

 \bibitem{Chernin1999}
A. D. Chernin, Mon. Not. R. Astron. Soc. 308, 321
(1999a).

\bibitem{Chernin1999b}
A. D. Chernin, Pis▓ma Astron. Zh. 25, 684 (1999b)
[Astron. Lett. 25, 591 (1999)].

\bibitem{CherninZasov2000}
A. D. Chernin, A. V. Zasov, V. P. Arkhipova, and
A. S. Kravtsova, Pis▓ma Astron. Zh. 26, 342 (2000)
[Astron. Lett. 26, 285 (2000)].

\bibitem{Chernin2001}
A. D. Chernin, A. S. Kravtsova, A. V. Zasov, and
V. P. Arkhipova, Astron. Zh. 78, 963 (2001a) [Astron.
Rep. 45, 841 (2001)].

\bibitem{CherninZasov2001}
A. D. Chernin, A. V. Zasov, V. P. Arkhipova, et al.,
ASP Conf. Ser. 230, 147 (2001b).

\bibitem{CherninZasov-2001!hexag-ring}
A. D. Chernin, A. V. Zasov, V. P. Arkhipova, et al.,
Astron. Astroph. Trans. 20, 139 (2001c).

\bibitem{CherninKovKor2006}
A. D. Chernin, V. V. Korolev, and V. V. Kovalenko,
Astrophys. Space Sci. Lib. 337, 321 (2006).

 \bibitem{Contopoulous1986}
G. Contopoulous and P. Grosbol, Astron. Astrophys.
155, 11 (1986).

 \bibitem{Cox-Gomez-2002!Analytical-Expressions-Spiral-Arm}
D. P. Cox and G. C. Gomez, Astrophys. J. Suppl. Ser.
142, 261 (2002).

 \bibitem{Dias-Lepine-2005!corot-MW}
W. S.Dias and J. R.D. Le? pine, Astrophys. J. 629, 825
(2005).

\bibitem{Dobbs-Bonnell-2006!Spurs-feathering}
C. L. Dobbs and I. A. Bonnell, Mon. Not. R. Astron.
Soc. 367, 873 (2006).

\bibitem{Efremov2009}
Yu. N. Efremov, Pis▓ma Astron. Zh. 35, 563 (2009)
[Astron. Lett. 35, 507 (2009)].

  \bibitem{Efremov-2010!M31}
Yu. N. Efremov,Mon. Not. R. Astron. Soc. 405, 1531
(2010).

   \bibitem{Efremov-2001!M31}
Yu. N. Efremov, Astron. Astroph. Trans. 20, 115
(2001).

\bibitem{Efremov2003}
Yu. N. Efremov and A. D. Chernin, Usp. Fiz. Nauk
173, 3 (2003) [Phys. Usp. 46, 1 (2003)].

 \bibitem{Elmegreen-1980!Spurs}
B. G. Elmegreen, Astrophys. J. 242, 528 (1980).

\bibitem{Elmegreen-etal-1998!NGC925}
B. G. Elmegreen, E. Wilcots, and D. J. Pisano, Astrophys.
J. Lett. 494, 37 (1998).

   \bibitem{Eremin-2010!Politex}
M. A. Eremin, A. V. Khoperskov, and S. A. Khoperskov,
Izv. VolGTU, Aktual. Probl. Upravl., Vychisl.
Tekhn. Informat. 13, 24 (2010).

 \bibitem{Gerhard-2010!Pattern-speeds-MW}
O. Gerhard, arXiv1003.2489 (2010).

 \bibitem{Grosbol-Dottori-2009!corot-N2997}
P. Grosbol and H. Dottori, Astron. Astrophys. 499, 21
(2009).

 \bibitem{Harten1983}
A. Harten, J. Comput. Phys. 49, 357 (1983).

 \bibitem{Khoperskov-2007}
A. V. Khoperskov, M. A. Eremin, M. A. Butenko,
and S. S. Khrapov, Vestn. Volgogr. Gos. Univ., Ser.
Matem. Fiz. 11, 105 (2007).

 \bibitem{Khoperskov-2003}
A.V. Khoperskov, S. S. Khrapov, and E. A. Nedugova,
Pis▓ma Astron. Zh. 29, 288 (2003) [Astron. Lett. 29,
246 (2003)].

 \bibitem{Khoperskov-etal-2010!AN}
A. Khoperskov, D. Bizyaev, N. Tiurina, et al., Astron.
Nach. 331, 731 (2010).

\bibitem{Kim-etal-2008!Spiral-Shocks-Thermal-Instability}
Ch.-G. Kim, W.-T. Kim, and E. C. Ostriker, Astrophys.
J. 681, 1148 (2008).

 \bibitem{Leda}
LEDA, HyperLeda, http://leda.univ-lyon1.fr/

 \bibitem{vanLeer1979}
B. van Leer, J. Comput. Phys. 32, 101 (1979).

\bibitem{Mishurov-etal-2002!chem-abundance-MW-corot}
Yu. N.Mishurov, J. R. D. Lepine, and I. A. Acharova,
Astrophys. J. Lett. 571, 113 (2002).

 \bibitem{Muraoka-etal-2009!spurs-M83}
K. Muraoka, K. Kohno, T. Tosaki, et al., Astrophys.
J. 706, 1213 (2009).

 \bibitem{Ned}
NED, NASA√IPAC Extragalactic Database≈NED,
http://ned.ipac.caltech.edu/.

 \bibitem{Patsis-etal-1997!Model-simul-gal}
P. A. Patsis, P. Grosbol, and N. Hiotelis, Astron.
Astrophys. 323, 762 (1997).


 \bibitem{Pohlen-Trujillo-2006!obs-gal}
M. Pohlen and I. Trujillo, Astron. Astrophys. 454, 759
(2006).

 \bibitem{Rautiainen-etal-2008!Corotat-SB-gal}
P. Rautiainen, H. Salo, and E. Laurikainen,Mon.Not.
R. Astron. Soc. 388, 1803 (2008).

 \bibitem{Scarano-etal-2010!corot-chem-dyn}
S. Scarano, J. Lepine, and M. Marcon-Uchida,
Mon. Not. R. Astron. Soc. (in press) arXiv:1012.5794
(2011).

  \bibitem{SDSS}
SDSS, NGC Galaxies in the SDSS Data Release 2
(DR2), http://www.sdss.org/dr2/.

\bibitem{Shetty-Ostriker-2006!Global-Modeling-Spur-Formation}
R. Shetty and E. C. Ostriker, Astrophys. J. 647, 997
(2006).

\bibitem{Shetty-Ostriker-2008!Cloud-Star-Formation-grav}
R. Shetty and E. C. Ostriker, Astrophys. J. 684, 978
(2008).

 \bibitem{Silchenko-Moiseev-2006!Nuclear-Rings}
O. K. Sil▓chenko and A.V.Moiseev,Astronom. J. 131,
1336 (2006).

  \bibitem{LaVigne-2006!Feathers-Spiral-Galaxies}
M. A. la Vigne, S. N. Vogel, and E. C. Ostriker,
Astrophys. J. 650, 818 (2006).

 \bibitem{Vorontsov1964}
B. A. Vorontsov-Vel▓yaminov, Astron. Zh. 41, 814
(1964) [Sov. Astron. 8, 649 (1964)].

\bibitem{Vorontsov1977}
B. A.Vorontsov-Vel▓yaminov, Extragalactic Astronomy
(Nauka, Moscow, 1977; Harwood Acad., 1987).

\bibitem{Wada-2008!Instabilities-Spiral-Shocks}
K.Wada, Astrophys. J. 675, 188 (2008).

\bibitem{Wada2001}
K. Wada and J. Koda, Publ. Astron. Soc. Jpn. 53,
1163 (2001).

\bibitem{Waller-etal-1997!polyg-obs-M101}
W. H. Waller, R. C. Bohlin, R. H. Cornett, et al.,
Astrophys. J. 481, 169 (1997).

\bibitem{Zasov-etal-2004!stat}
A. V. Zasov, A. V. Khoperskov, and N. V. Tyurina,
Pis▓ma Astron. Zh. 30, 653 (2004) [Astron. Lett. 30,
593 (2004)].

 \bibitem{Zhang-Buta-2007!Corotation-Radii-Spiral}
X. Zhang and R. J. Buta, Astron. J. 133, 2584 (2007).

\bibitem{Zhang-Buta-2010!Morphological-Transformation-Galaxies}
X. Zhang and R. J. Buta, arXiv:1012.0277 (2010).

}

%$$$$$$$$$$$$$$$$$$$$$$$$$$$$$$$$$$$$


\end{thebibliography}
\end{document}